\documentclass[twocolumn]{article}

\usepackage{arxiv}
\usepackage{float}
\usepackage{subcaption}  % For subfigures with individual captions

\usepackage[utf8]{inputenc} % allow utf-8 input
\usepackage[T1]{fontenc}    % use 8-bit T1 fonts
\usepackage{hyperref}       % hyperlinks
\usepackage{url}            % simple URL typesetting
\usepackage{booktabs}       % professional-quality tables
\usepackage{amsfonts}       % blackboard math symbols
\usepackage{nicefrac}       % compact symbols for 1/2, etc.
\usepackage{microtype}      % microtypography
\usepackage{lipsum}		% Can be removed after putting your text content
\usepackage{graphicx}
\usepackage[numbers]{natbib}
\usepackage{doi}
\usepackage{amsmath}
\usepackage{caption}

\title{Impact of protein corona morphology on nanoparticle diffusion in biological fluids: insights from a mesoscale approach}

\author{ \href{https://orcid.org/0000-0002-3983-1587}{\includegraphics[scale=0.06]{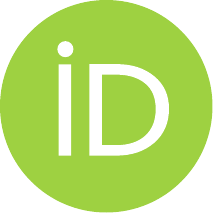}\hspace{1mm}Beatrice Cipriani*}\\%\thanks{Supplementary Information available: plasma and corona proteins' details, derivation of Moment of Inertia, Brownian Dynamics computational details.} \\
	FOCAS Research Institute \\ Technological University of Dublin \\ City Campus, Dublin 8, Ireland \\
	\texttt{d20127863@mytudublin.ie} \\
	\And
	\href{https://orcid.org/0000-0003-1083-6234}{\includegraphics[scale=0.06]{orcid.pdf}\hspace{1mm}Hender Lopez*} \\
	School of Physics, Clinical and Optometric Sciences \\ Technological University of Dublin \\ Grangegorman D07 ADY7, Ireland \\
	\texttt{hender.lopezsilva@tudublin.ie} \\
}

\renewcommand{\shorttitle}{Impact of protein corona morphology on nanoparticle diffusion}
\hypersetup{
pdftitle={A template for the arxiv style},
pdfsubject={q-bio.NC, q-bio.QM},
pdfauthor={Beatrice Cipriani, Hender Lopez},
pdfkeywords={soft matter, protein corona, macromolecular crowding, particle mobility, effective sphere models},
}

\begin{document}
\onecolumn
\maketitle

\begin{abstract}
Nanoparticles (NPs) demonstrate considerable potential in medical applications, including targeted drug delivery and diagnostic probes. However, their efficacy depends on their ability to navigate through the complex biological environments inside living organisms. In such environments, NPs interact with a dense mixture of biomolecules, which can reduce their mobility and hinder diffusion. Understanding the factors influencing NP diffusion in these environments is key to improving nanomedicine design and predicting toxicological effects. In this study, we propose a computational approach to model NP diffusion in crowded environments. We introduce a mesoscale model that accounts for the combined effects of the Protein Corona (PC) and the crowded medium on NP movement. By including volume-exclusion interactions and modelling the PC both explicitly and implicitly, we identify key macromolecular descriptors that affect NP diffusion. Our results show that the morphology of the PC can significantly affect the diffusion of NPs, and the role of the occupied volume fraction and the size ratio between tracers and crowders are analysed. The results also show that approximating large macromolecular assemblies with a hydrodynamic single-sphere model leads to inexact diffusion estimates. To overcome the limitations of single-sphere representations, a strategy for an accurate parametrization of NP-PC systems using a single-sphere model is presented.
\end{abstract}

% keywords can be removed

\begin{center}
\includegraphics[width=0.65\linewidth]{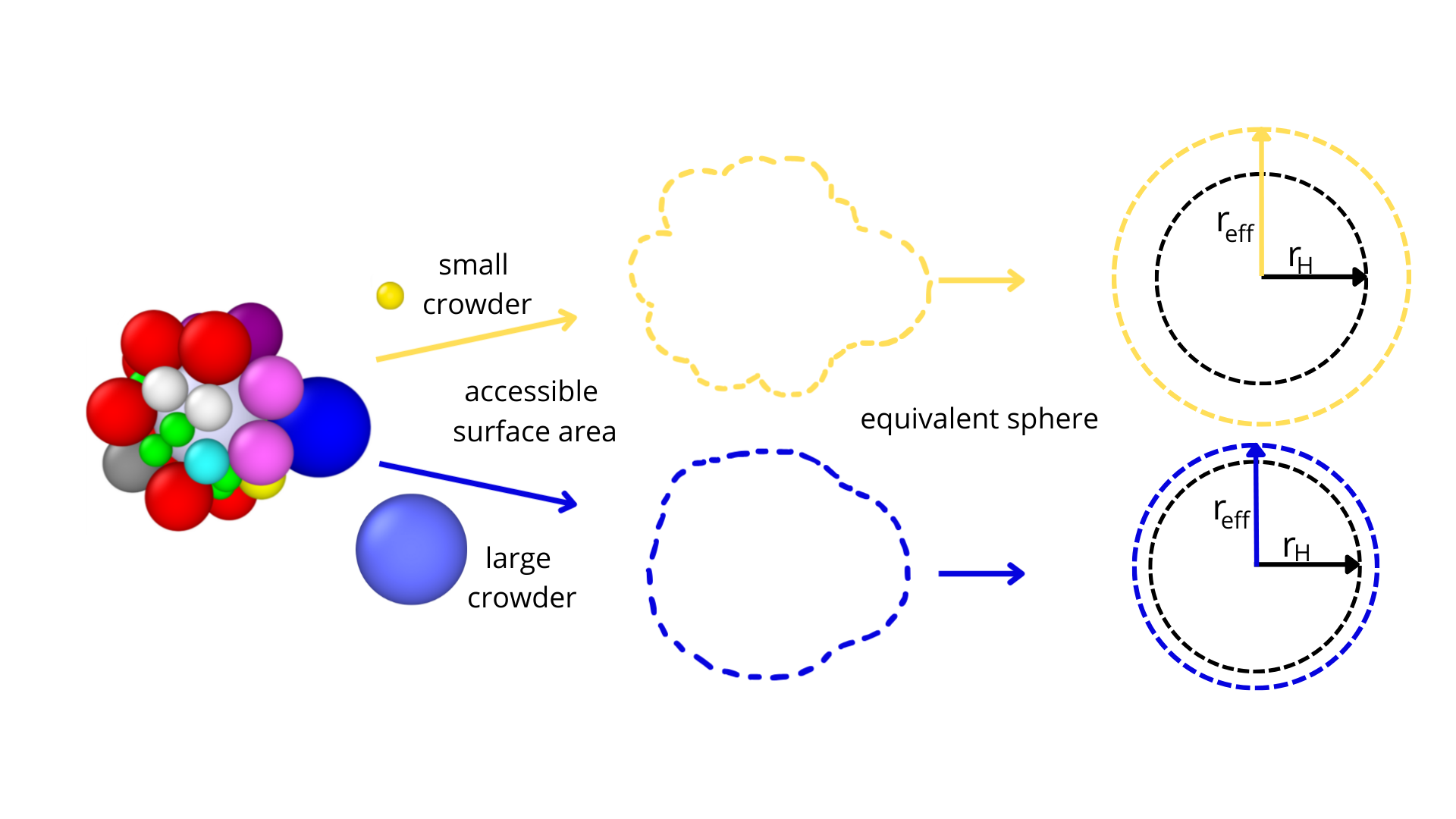}
\captionof*{figure}{\textbf{Graphical abstract}}
\end{center}

% ----- Switch to two-column layout for main text -----
\twocolumn

\section{Introduction}
The potential use of nanoparticles (NPs) in biological, pharmaceutical, chemical, and medical fields has produced an interdisciplinary research field known as nanomedicine \citep{Riehemann2009}. It is worth nothing that effective nanomedical applications rely on the NP’s ability to target and navigate specific structures within living organisms. In this way, assessing the mobility of NPs is a critical factor that can influence potential applications, requiring the evaluation of several key aspects. First, NP's shape and surface composition are known to affect their bioactivity \citep{Zhao2011}. Nowadays, NPs can be designed with a multitude of shapes, such as spheres, flowers, or stars, and their surface morphology is often engineered to include features like spikes, virus-  or raspberry-like patterns \citep{Yu2019, Li2018, Tian2015, Zhang2022, Cai2022}. Additionally, when NPs interact with biological fluids, they get immediately coated by a biomolecule rich layer known as protein corona (PC). Its composition depends on several factors such as the NP's size, shape, material, and the biological medium content \citep{Cedervall2007, Monopoli2011, Kopac2021}. However, it is clear that its formation leads to morphological changes, resulting in anisotropic structures with patchy patterns on the surface \citep{Sheibani, Picco2021, Cao2021, Bergese2025}. As a result, NPs with a PC are unlikely to maintain an ideal, smooth spherical shape. This transformation impacts the NP’s mobility and transport properties by altering its size, shape, interaction dynamics and overall its biological identity, thus influencing their bio-distribution and bio-availability \citep{Qin2020, Tenzer2013, Etoc2018, Ritz2015}.
Another factor to consider is that the biological environments where the transport happens introduces additional complexity to the dynamics of the NPs due to macromolecular crowding. Biological fluids, particularly within cells, are densely populated with proteins and other macromolecules that can occupy up to 40$\%$ of the total volume. Furthermore, these crowded environments are often polydisperse, complicating the analysis and study of transport properties. To understand the dynamics of NPs in such conditions, we can draw connections with the well-studied behaviour of colloids, an approach that has also been applied to protein solutions \citep{Stradner2020}. The colloid-protein analogy has been extensively explored in soft matter physics, as globular proteins often exhibit colloidal-like behaviour in solution. Over the years, this homology has provided insights into protein phase diagrams, crystallization, self-assembly, aggregation, and diffusion \citep{Stradner2004, Foffi2014, Ando2010}. It is well known that colloidal systems at high concentrations display reduced diffusion rates due to particle interactions, including hydrodynamic effects, crowding, and collisions \citep{Blanco2017, Roosen-Runge2011}. Scaling theories describing concentrated monodisperse soft sphere colloids and polymers have effectively captured these phenomena \citep{Tokuyama1995, Phillies1987, Langevin1978, Cai2011}.
However, for non-spherical particles, the application of these theories is by no mean straightforward. Surface roughness and particle shape influence diffusion rates, with rough particles diffusing more slowly than smooth ones near walls in pure solvents \citep{Lang2021}. Two recent computational studies have emphasized the role of morphology in NP diffusion \citep{Moreno-Chaparro2023, Wani2024}. The first study examined how the distribution, size, and morphology of functional groups on spherical NPs influence their translational diffusion both at infinite dilution and near rigid walls. The findings revealed that the transport properties of functionalized NPs are notably affected by the morphology of the attached groups. In the second study, it was shown that NPs with different anisotropic shapes display unique self-diffusion and sedimentation behaviours, with a stronger dependence on volume fraction compared to spheres. Moreover, it has been proved how in crowded media, interactions among rough colloids can lead to rotational arrest and phenomena such as second glass transition in both translational and rotational diffusion, which are absent for systems with smooth particles \citep{Ilhan2022}.
While colloid-based models have successfully described the dynamics of some proteins under specific crowded conditions \citep{Blanco2017, Grimaldo2019,Ando2010}, they are typically restricted to spherical and isotropic particles in monodisperse solutions. More complex modelling approaches are often required to address anisotropic objects \citep{Stradner2020, Hirschmann2023, Roosen-Runge2011}. For example, it has been demonstrated that anisotropic shapes and interactions significantly influence protein diffusion, particularly at high concentrations \citep{Balbo2013, Gulotta2024, Bucciarelli2016}. All of these findings suggest that understanding NP diffusion in biological fluids requires integrating macromolecular crowding and PC-induced morphological features into existing models.
NP diffusion plays a crucial role in determining the \textit{in situ} properties of the PC, particularly its thickness. Hydrodynamic size measurements, often obtained through fluorescence-based techniques \citep{Felix2013, Skora2020, Kwon2020, Kuhn2011, Babayekhorasani2016}, rely on diffusion models that typically assume spherical particles. This simplification, commonly adopted in both experimental and simulation studies to enable more tractable modelling and analysis, treats complex structures as effective spheres \citep{Roosen-runge2021}. These are typically defined by an equivalent hydrodynamic radius, derived from the diffusion coefficient at infinite dilution of the non-spherical body. This approach is widely used to estimate PC thickness in biological fluids \citep{Cui2014, Nikitin2021}. Since these methods are grounded in diffusion theory, an accurate understanding of nanoparticle mobility is essential for reliable measurements. \textit{In situ} methods are generally preferred over \textit{ex situ} approaches, as the latter can be significantly affected by separation techniques \citep{Kato2010, Bohmert2020}.

In this context, this work aims to model the diffusion of NPs in crowded, polydisperse environments. For this task, we developed a computational model to simulate NP dynamics in protein-rich media, accounting for macromolecular crowding and the presences of a PC on spherical NPs. We adopt a Coarse-Grained Meso-Scale (CG-MS) approach that ensures computational feasibility over the timescales relevant for the long-time diffusion. Here, proteins are modelled as spheres, and NPs with a PC are treated as a single rigid objects using a raspberry-like model, a representation commonly used in computational studies of NPs with a PC \citep{Rouse2021, Tavanti2019, MosaddeghiAmini2023}.
With this model, we then quantify how the PC properties and macromolecular crowding jointly affect NP diffusion and evaluate how the combination of these two factors might influence the effective single-sphere approximation. To achieve this, our analysis explores variations in PC morphology, spatial organization, and composition to assess their impact on diffusion. We compare explicit against implicit PC representation, with the aim of determining the conditions under which the equivalent single-sphere approximation is valid. Finally, we evaluated the role of the medium composition (monodisperse versus polydisperse crowders) and of the occupied volume fraction in NP dynamics in biological environments.

\section{Methods}

NPs typically diffuse through biological fluids under crowded molecular conditions, often coated by a PC layer. Simulating such systems at the atomistic resolution level can be computationally expensive. For instance, considering a 20-nm diameter spherical NP in a box of $110\times110\times110$ nm$^3$ at a crowding volume fraction $\phi=0.3$ would require simulating between 1,000-10,000 proteins over time scales relevant to NP diffusion - usually on the order of several microseconds or even milliseconds. Employing full-atomistic or even moderately coarse-grained models to reach these timescales for a few hundreds of proteins (corresponding to $\approx$50,000-100,000 residues) would result in a massive investment of computational resources. Moreover, when simulating a highly polydisperse system, we must account for multiple replicas to obtain robust statistics.
In this work, to address these challenges, we adopted a mesoscale representation of the proteins in solution and within the PC, wherein each protein is represented by a single sphere. This approach preserves the tracer's morphological properties, allowing us to assess their contribution to diffusion, while with spherical crowders we can still evaluate different size contribution and concentration effects through volume exclusion interactions. The size of each sphere is set to the hydrodynamic radius of the protein computed in pure solvent. Using this approach enables efficient exploration of the NP's diffusion within complex biological environments while balancing computational costs and system comparability.

\subsection{Nanoparticle and Protein Corona (NP-PC) models}
\label{sec:NP-PC model}
In this work, each NP is modelled as a sphere of 20 nm in diameter while the PC is represented by spheres of different sizes (each size representing a protein type) attached to the NP. These sizes were set to the hydrodynamic radius ($r_\mathrm{H}$) of the proteins obtained by computing the translational diffusion coefficient ($D_\mathrm{t_0}$) of each protein type at infinite dilution (\textit{i.e.} in pure solvent) using HYDROPRO software (version 10) \cite{Ortega2011} and then using the Stokes-Einstein relationship to calculate $r_\mathrm{H}$. The atomic coordinates for the atoms of the proteins used for these calculations were obtained from the Protein Data Bank \cite{rcsb} when available, and alternatively from the Alphafold repository \cite{alphafold} (see Table~S1 for details). The viscosity and density of the solvent were set to 0.01 poise and 1 g/ml, respectively, while a temperature of 293 K was used in all calculations. The number of binding sites available on the surface of the NP, and therefore the number of proteins that can be forming the PC, was calculated for each protein type using the approach developed by Rouse and Lobaskin \cite{Rouse2021}. This method accounts for the steric occupancy of every protein, and is developed as follows. First, the projected binding area of each protein that constitutes the PC is calculated with the formula,
	\begin{equation}
		a_\mathrm{i} = 2\pi r_\mathrm{\text{NP}}^2\left[1-\sqrt{1-\left(\frac{r_\mathrm{i}}{r_\mathrm{i}+r_\mathrm{\text{NP}}}\right)^2}\right],
	\label{eq:PC_model1}
	\end{equation}
	where $r_\mathrm{NP}$ is the NP's radius, and $r_\mathrm{i}$ is the radius of the i-th protein. The total surface of the NP covered by all proteins of a certain type is calculated as,
	\begin{equation}
		A_\mathrm{i} = [C_\mathrm{i}] \cdot 4\pi r_\mathrm{\text{NP}}^2 ,
	\label{eq:PC_model2}
	\end{equation}
where $C_\mathrm{i}$ is the relative abundance of each corona protein type. This parameter can be arbitrary set to design a desired corona composition, or derived from experimental data of real PC. The number of binding sites available for each protein type is then given by $s_\mathrm{i} = A_\mathrm{i}/a_\mathrm{i}$. In this study, we modelled only mono-layered PCs, keeping the total surface coverage of the NPs between 85-90$\%$ for all the simulated systems.
For a given PC composition, the spheres that constitute the PC are randomly distributed on the surface of the NP. These proteins are modelled as hard spheres, avoiding any overlap. As different configurations of the same PC might have different effects on the diffusivity of the NP-PC complex, we generate multiple layouts of each PC, \textit{i.e.} the same composition of the PC but with the different arrangements of the proteins. In practice this is done as follows; prior to the main simulations, a randomly PC arrangement is generated. Then we perform a brief Brownian Dynamics (BD) simulation involving the NP and the corona proteins alone. At this stage, the proteins are allowed to diffuse on the surface, and in this way we are able to model the same protein composition but with different layouts around the NP. Each frame of the BD trajectory corresponds to a unique PC configuration around the NP. Afterward, during the production simulation, the NP-PC is treated as rigid-body (RB), \textit{i.e.} we assumed that the PC is irreversibly bound to the NP once formed \cite{Milani2012} and that their layout and composition do not change in time. To enable this, both translational and rotational drag coefficients, as well as the moment of inertia (MOI) of the RB, had to be defined. The translational diffusion coefficient ($D_\mathrm{t_0}$) and the $3\times3$ rotational diffusion matrix at infinite dilution were computed for the chosen RB configuration using HYDRO++ \cite{DeLaTorre2007}. The corresponding translational drag coefficient and rotational drag coefficient tensor were then obtained by taking the reciprocals of these values. The physical properties of the solvent used in these calculations were the same as those reported above for the plasma proteins. Furthermore, the MOI of the RB was calculated, and the principal axes of inertia were aligned with the simulation box axes (x, y, z).

\begin{figure}[h]
\centering
  \includegraphics[width=8.1cm]{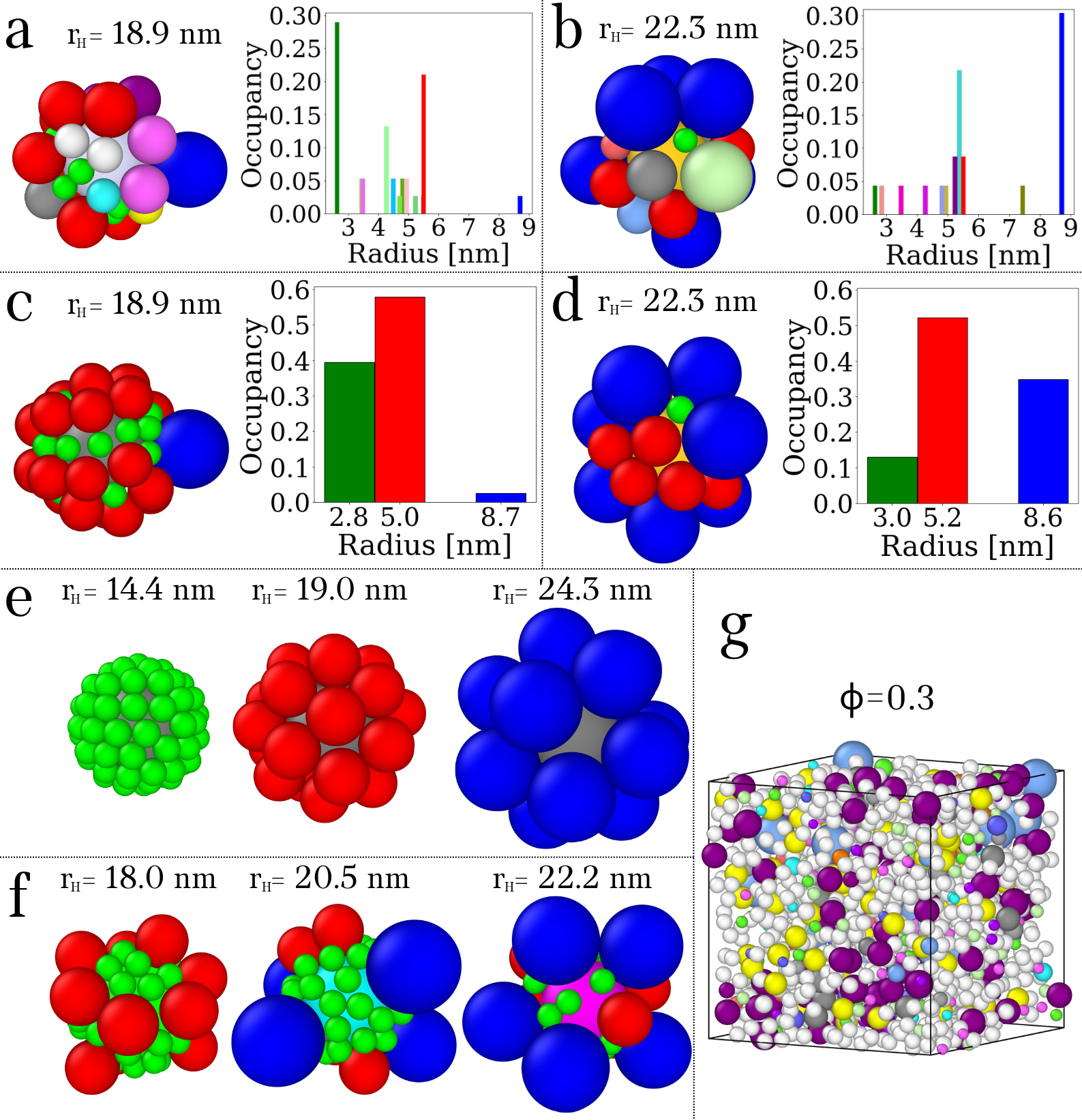}
	\caption{a-b: representation of the NP-PC complex and the protein size distribution of the corona for the models P1 (a) and P2 (b). c-d: representation of the NP-PC complex and the protein size distribution of the corona for the simplified corona models P1-S (c) and P2-S (d). e: from left to right, graphical representation of MS NP, MM NP and ML NP models. f: the three final systems investigated in this work were designed with different molar ratios of small (green), medium (red) and large (blue) proteins in the corona, resulting in three qualitatively different shapes. From left to right, the relative concentration ratio of the corona proteins is 1:1:0 (P3), 2:1:1 (P4), and 1:1:2 (P5). (g) Example of a simulation box containing plasma proteins at a total volume fraction of $\phi_\mathrm{tot} = 0.3$. The protein types and their relative concentrations are based on experimental plasma composition data.}
	\label{fig:methods}
\end{figure}

The first two systems modelled are 20-nm NPs with polydisperse plasma PCs. The two proteomic compositions of the PC were selected from experimental data, specifically from 20 nm Silver and Gold NPs citrate coated (Ag-CIT and Au-CIT NPs) as reported in Ref\cite{Lai2017} (See Table~S2-S3 for PCs composition). We will refer to these systems as P1 and P2 NPs. Figures~\ref{fig:methods} a-b show the relative occupancy of each protein in the corona and a rendering of the final system. 

As mentioned above, different configurations of the same PC were generated. For both P1 and P2 NPs, 1000 different layouts of the PC were constructed. For each of these configurations of the NP with a PC, the hydrodynamic radius of the NP-PC was derived from the computed $D_\mathrm{t_0}$ using the Stokes-Einstein relationship. To investigate the role of PC's layout on the long-time diffusivity, from this set of 1000 configurations, we selected four for the production BD simulations in crowded medium. The four configurations were selected based on the following criteria: a NP-PC with the modal hydrodynamic size ($t_1$), a NP-PC with the largest hydrodynamic radius ($t_2$), a NP-PC with the smallest hydrodynamic radius ($t_3$), and a randomly selected NP-PC ($t_4$), as shown in Figures~S1-S2. Additionally, we derived two simplified models, referred to as P1-S and P2-S, from the original P1 and P2 NP-PC systems. In these simplified models, the PC is composed of only three distinct protein sizes (Figure~\ref{fig:methods}c-d), in contrast to the full distribution of protein sizes present in the original P1 and P2 models (Figure~\ref{fig:methods}a-b). These three sizes were obtained by discretising the original protein distributions into histograms, grouping proteins with similar hydrodynamic radii. For the P1-S model, the protein sizes were grouped into three representative radii: 2.8~nm, 5.0~nm, and 8.7~nm. The number of size types was chosen to preserve the overall hydrodynamic radius of the NP-PC, such that the resulting $r_\mathrm{H}$ of the P1-S matches that of the original P1 NP-PC of 18.9~nm. Similarly, for the P2-S model, the simplified composition includes proteins with radii of 3.0~nm, 5.2~nm, and 8.6~nm, also preserving the same $r_\mathrm{H}$ as the original P2 NP-PC of 22.3~nm. This reduction in complexity was implemented to test whether a simplified representation of the PC, in which the detailed size distribution is reduced but the overall hydrodynamic size and morphology are preserved, has any impact on tracer diffusion.

\begin{table}[h]
\small
  \caption{All NP-PC systems investigated in this study, with full names, ID and hydrodynamic size. The asterisk (*) indicates that for these systems, the equivalent h-SS model has been simulated.}
  \label{tab:systems}
  \begin{tabular*}{8.1cm}{@{\extracolsep{\fill}}lll}
    \hline
    System & ID & $r_\mathrm{H}$\\ && [nm]\\
    \hline
    Ag-CIT NP with Polydisperse PC & P1 $t_1$ $*$ & 18.9 \\
      & P1 $t_2$ & 19.0 \\
      & P1 $t_3$ & 18.7\\
      & P1 $t_4$ & 18.8\\
    Au-CIT NP with Polydisperse PC & P2 $t_1$ $*$ & 22.3\\
      & P2 $t_2$ & 22.5 \\
      & P2 $t_3$ & 22.1 \\
      & P2 $t_4$ & 22.3 \\
    P1 NP with Simplified PC & P1-S & 18.9 \\
    P2 NP with Simplified PC & P2-S & 22.3 \\
    NP with Polydisperse PC (1:1:0 protein ratio) & P3 & 18.0 \\
    NP with Polydisperse PC (2:1:1 protein ratio) & P4 & 20.5 \\
    NP with Polydisperse PC (1:1:2 protein ratio) & P5 & 22.2 \\
    NP with Monodisperse PC (small) & MS $*$ & 14.5\\
    NP with Monodisperse PC (medium) & MM $*$ & 19.0 \\
    NP with Monodisperse PC (large) & ML $*$ & 24.3 \\
    \hline
  \end{tabular*}
\end{table}

To assess whether the diffusion of the NP can be influenced by the polydispersity of the PC, we modelled three more 20 nm NPs with monodisperse PCs, \textit{i.e.} made only of one kind of protein (Figure~\ref{fig:methods}-e). As protein types we chose three types: small size proteins (spheres of radius 2.6 nm), medium size proteins (spheres radius 5.5 nm) and large proteins (spheres of radius 8.7 nm). These sizes were selected as they are commonly found in both the P1 and P2 coronas (see Figure~\ref{fig:methods} a-b). We will refer to these systems as Monodisperse Small (MS), Monodisperse Medium (MM), and Monodisperse Large (ML) coronas. 
For these five systems (P1, P2, MS, MM and ML NPs) we perform BD simulations both by treating them as rigid-bodies (RBs) and by modelling the NP-PC complex as a single sphere of size determined by the NP-PC hydrodynamic size (equivalent hydrodynamic single sphere, h-SS). In this way, instead of representing the NP-PC system by the use of a multi-beads RB as shown in Figure~\ref{fig:methods}, we now represent the NP-PC complex as a single sphere of which its size is derived using the Stokes-Einstein formula from the $D_\mathrm{t_0}$ calculated with HYDRO++ from the RB structure. Thus, we are able to simulate the PC both explicitly and implicitly, evaluating the validity of the h-SS approximation on NPs diffusion. Finally, we modelled three more 20 nm NPs with a simplified polydisperse PCs, \textit{i.e.} using only three protein types. The corona compositions of these three systems were not based on experimental data, but were artificially designed to reproduce three different morphologies - from a more compact homogeneous-shaped NP to a more rough and irregularly-shaped one (Figure~\ref{fig:methods}-f). We will refer to these as P3, P4, and P5. A summary of all NP-PC systems is presented in Table~\ref{tab:systems}.

\subsection{Medium model}
\label{sec:medium model}

In this study we simulated NPs diffusing in different media at different concentrations. We modelled both polydisperse and monodisperse suspensions. For the polydisperse suspension, we simulated human plasma, a common incubation medium for in vitro experiments. Human plasma is a complex suspension of biomolecules containing over 3,000 identified proteins but the 20 most abundant represent approximately 98$\%$ of the total protein mass of plasma \cite{Hortin2006}. To provide a comprehensive representation of polydispersity of the plasma proteome, the plasma model simulated was composed of the 29 most abundant proteins in plasma. The hydrodynamic properties for each protein at infinite dilution were calculated from the full-atomistic protein structures using HYDROPRO, with the same parameters as reported in Section~\ref{sec:NP-PC model}. After, each protein in the medium was represented by a single bead of the size of its $r_\mathrm{H}$. Physical properties of the plasma proteins are reported in Table~S1. Finally, the polydispersity index of the medium was calculated as $\alpha = \sigma/\langle r_\mathrm{H}\rangle$, where $\sigma$ is the standard deviation of the hydrodynamic sizes in solution and $\langle r_\mathrm{H}\rangle$ is the mean of the size distribution.

In the monodisperse suspensions, a NP-PC was allowed to diffuse within a simulation box containing only one crowder type (corresponding to $\alpha=1$). The crowder sizes considered in this study range from 2.1 nm to 25 nm, covering a broad spectrum of possible crowding conditions. The selected sizes include: $2.1$ nm, representing the smallest protein in the polydisperse medium; $3.5$ nm, corresponding to the most abundant protein in the polydisperse medium; $3.7$ nm, which represents the average size of all proteins in the polydisperse medium; $3.9$ nm, obtained using the effective radius formula $r_\mathrm{eff} = \sqrt[3]{\langle r_\mathrm{i}^3\rangle}$ from~\cite{Grimaldo2019}, where angle brackets denote an average size over the entire distribution of proteins in the human plasma model (this value should provide an effective monodisperse equivalent for the polydisperse medium); $8.7$ nm, corresponding to the largest protein in the polydisperse medium; $18.9$ nm, matching the hydrodynamic size of the tracer in solution (P1 NP); $25$ nm, included to extend the range of crowding conditions explored, specifically to scenarios where the crowders are larger than the tracer in solution. Table~\ref{tab:medium} summarizes all the media tested, detailing their composition, average size, and volume occupancy.

\begin{table}[h]
\small
  \caption{\ All the media modelled and investigated in this study. $\langle r_\mathrm{cr} \rangle$ indicates the average hydrodynamic size of the crowders in solution.}
  \label{tab:medium}
  \begin{tabular*}{8.1cm}{@{\extracolsep{\fill}}l l l}
    \hline
    Composition & $\langle r_\mathrm{cr} \rangle$ [nm]& $\phi$ \\
    \hline
    Polydisperse plasma & 3.7 & 0.05, 0.1, 0.2, 0.3, 0.4, 0.5 \\
    Mono-crowded & 2.1 & 0.3 \\
    Mono-crowded & 3.5 & 0.005, 0.025, 0.05, 0.1, \\ &&0.2, 0.3, 0.4, 0.5 \\
    Mono-crowded & 3.7 & 0.3 \\
   Mono-crowded & 3.9 & 0.3 \\
   Mono-crowded & 4.1 & 0.3 \\
   Mono-crowded & 4.3 & 0.3 \\
   Mono-crowded & 8.7 & 0.05, 0.1, 0.2, 0.3, 0.4, 0.5 \\
   Mono-crowded & 18.9 & 0.3 \\
   Mono-crowded & 25.0 & 0.3 \\
    \hline
  \end{tabular*}
\end{table}

\subsection{Brownian Dynamics (BD)}
To investigate the diffusion of NPs with a PC layer under macromolecular crowding conditions, we performed overdamped BD simulations using the HOOMD-blue software package\cite{Anderson2020}. For the NP-PC RB, an anisotropic integrator was used, ensuring that the correct MOI were defined. BD simulations were performed in a cubic box with periodic boundary conditions. The box size for each system was determined to ensure it was at least three times the size of the tracer, with the NP-PC volume fraction, $\phi_\mathrm{NP}$, held constant across simulations within the same medium. For example, in the plasma medium with $\phi_\mathrm{tot} = 0.3$, the NP-PC volume fraction was set to $\phi_\mathrm{NP} = 0.015$. A total of 1086 proteins were initially modelled for the plasma medium, with their quantities determined based on the experimental molarities provided in Table~S1. The number of each protein species was set according to its respective molarity, preserving the experimental concentration ratios in the plasma medium. The volume occupied by this set of proteins was calculated as

\begin{equation}
V_\mathrm{plasma} = \sum_i [C_i] \cdot \frac{4}{3} \pi r_i^3,
\end{equation}

where $C_i$ is the molarity (or total number) of protein $i$ and $r_i$ is the hydrodynamic radius of that protein. The volume occupied by the crowders (proteins) in the box, needed to satisfy the chosen total volume fraction $\phi_\mathrm{tot}$, was then calculated as

\begin{equation}
V_\mathrm{crowders} = V_\mathrm{box} \cdot (\phi_\mathrm{tot} - \phi_\mathrm{NP}) = V_\mathrm{box} \cdot \phi_\mathrm{crowders}.
\end{equation}

Since $\phi_\mathrm{crowders}$ and $\phi_\mathrm{plasma}$ may differ, a scaling factor $f$ was introduced to adjust the number of proteins while maintaining their relative concentrations. The scaling factor was calculated as

\begin{equation}
f = \frac{\phi_\mathrm{crowders}}{\phi_\mathrm{plasma}}.
\end{equation}

Finally, the number of protein type $i$ in the simulation box was determined as

\begin{equation}
N_i = f \cdot [C_i].
\end{equation}

This approach ensures that the number of proteins in solution can be adjusted according to their relative concentrations, while respecting the constraints imposed by an appropriate box size and the desired volume fraction.

HOOMD-blue employs a self-consistent system of units \cite{Anderson2020}. The system temperature was set to 293 K, with energy units defined such that $k_BT=1$ where $k_B$ is Boltzmann constant. To establish a realistic time scale, we set the simulation length unit $\sigma=1$ nm and matched the translational diffusion coefficient $D_\mathrm{t_0}$ of both the proteins in solution and the tracer under dilute conditions to the values computed using HYDROPRO and HYDRO++, respectively \cite{DeLaTorre2007, Ortega2011}. This ensured that the reduced time unit was defined as $\tau = 1$ $\mu$s. For each system, between 5 and 50 independent systems were simulated, followed by production runs lasting up to 1250 \(\tau\). To avoid particle overlaps in dense systems, equilibration was performed by initially simulating a larger box and gradually compressing it to the target packing fraction. A fixed integration time step of $dt = 5 \times 10^{-5} \tau$ was used throughout the simulations.

During the BD simulations, one single NP-PC (RB or h-SS representation) is let to diffuse through the chosen medium. All the proteins and tracer in solution only interact via Weeks–Chandler–Anderson (WCA) potential \cite{WCA}, as per the formula:

\begin{equation}
V_\mathrm{WCA}(r) =
\begin{cases}
4\epsilon \left[\left(\dfrac{\sigma_\mathrm{ij}}{r_\mathrm{ij}}\right)^{12}
- \left(\dfrac{\sigma_\mathrm{ij}}{r_\mathrm{ij}}\right)^6 \right],
& r_\mathrm{ij} \le 2^{1/6}\sigma_\mathrm{ij} \\[6pt]
0, & r_\mathrm{ij} > 2^{1/6}\sigma_\mathrm{ij}
\end{cases}
\end{equation}

with $\epsilon$ being the depth of the potential, $r_\mathrm{ij}$ the distance between the centres of two interacting particles. The term $\sigma_\mathrm{ij}$ is the characteristic length at which the interaction between particles is zero. Here it is defined as $\sigma_\mathrm{ij} = (\sigma_\mathrm{i} + \sigma_\mathrm{j})/2$. The strength of repulsion was set as $\epsilon=2k_\mathrm{B}T$.

\subsection{Calculation of the translational diffusion coefficient}
For the NP-PC complexes and proteins in solution we calculated the ensemble averaged mean squared displacement (MSD) of the each particle type over time. For better statistics, we used the so called ``window method'', more specifically the implementation available in Freud's library \cite{Dice2019}. Then, from a linear fit of the  MSD vs. time curve, the translational diffusion was obtained from: 
\begin{equation}
	D_\mathrm{t} = \frac{m}{6},
\end{equation}
where $m$ is the slope of the linear fit. For the NP-PC types, the MSD is computed on the Centre of Mass of the RB. Care was taken to ensure that the slope was calculated in the long-time diffusive regime as detailed in Figure~S3-S8. 

\section{Results and discussion}
% Our results for the long time diffusion are presented in this section. First, we discuss the effects of the PC on the long-time diffusion of 10 different NP-PC systems (see Table~\ref{tab:systems} for details). Our analysis focuses on how various PC properties might impact diffusivity. Next, we explore the rigid-body (RB) and single-sphere (SS) models and their implications for describing macromolecular diffusion. All systems diffuse in a polydisperse plasma medium ($\#0$ in Table~\ref{tab:medium} with a fixed total volume fraction of $\phi = 0.3$ (Figure~\ref{fig:methods}-G).

% Finally, we analyse the effects of different macromolecular crowding conditions by simulating diffusion in a polydisperse medium across several volume fractions. We also investigate how the mobility of a representative NP-PC changes when transitioning from the polydisperse plasma suspension to media containing only a single type of crowder.

\subsection{Effects of the PC}

\begin{figure}[h]
\centering
  \includegraphics[width=8.1cm]{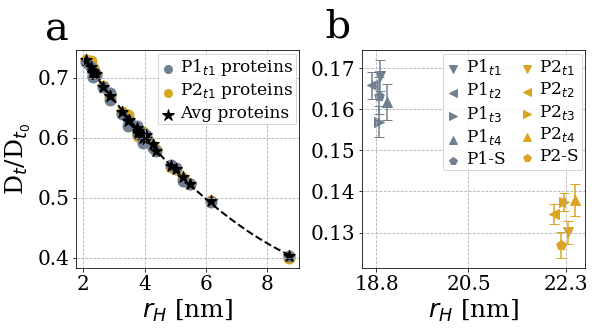}
	\caption{(a) Normalized, translational diffusivities for all the proteins in solution in the P1$_\mathrm{t1}$ (grey circles and P2$_\mathrm{t1}$ (gold circles) systems. Black stars indicate the protein diffusivities averaged over all the 10 systems discussed here. (b) Normalized, translational diffusivities for all the spatial arrangements (4+4) of P1 and P2 NPs (grey and gold triangles, respectively) and simplified representations (pentagons), with error bars. }
	\label{fig:results0}
\end{figure}

In this section, we focus on the effects of the PC on the long-time diffusion (see Table~\ref{tab:systems} for details on the systems simulated). In particular, we analysed how various PC properties might impact the diffusivity of the NP-PC complex. 
%The first results we present are those of all the P1 and P2 NPs models ($t_1, t_2, t_3, t_4$ and Simplified). 
Figure~\ref{fig:results0}-a reports the normalised long-time translational diffusion coefficients as a function of the hydrodynamic radius for the crowders in the medium, \textit{i.e.} plasma proteins for the systems P1$_\mathrm{t_1}$ and  P2$_\mathrm{t_1}$. Figure~\ref{fig:results0}-a also reports the average values of the diffusivities of the proteins in the medium across all 10 simulations (4 variations for each of the P1 and P2 NPs models and their Simplified versions), and comparing these values to the one for P1$_\mathrm{t_1}$ and P2$_\mathrm{t_1}$ indicates that there is very small variability between all simulated systems. A similar trend was noted for the remaining eight models. These findings suggest that the diffusion of proteins in suspension is not significantly affected by the specific type of NP-PC complex, at least under the conditions tested, where only a single NP-PC occupies a small volume fraction of the system. On the other hand, regarding the NPs, the results for the P1 and P2 NPs are depicted in Figure~\ref{fig:results0}-b. For each system, the four layouts ($t_1$, $t_2$, $t_3$, $t_4$) and the simplified PC (P1-S and P2-S) lead to nearly indistinguishable $D_\mathrm{t}$ estimates (within the error bars). These results suggest that simulating a single configuration of the PC is sufficient for accurately capturing the dynamics of the NP-PC complex. Also, reducing the number of protein types in the corona does not result in a significant loss of information, as long as the overall morphology is preserved. In this way, we establish that the specific layout of proteins in the corona does not heavily influence their overall mobility. Therefore, to simplify the following analysis we will only present the $t_1$ arrangement, without further discussion of the other cases ($t_2$, $t_3$, $t_4$ and Simplified).

\begin{figure}[h]
\centering
  \includegraphics[width=8.1cm]{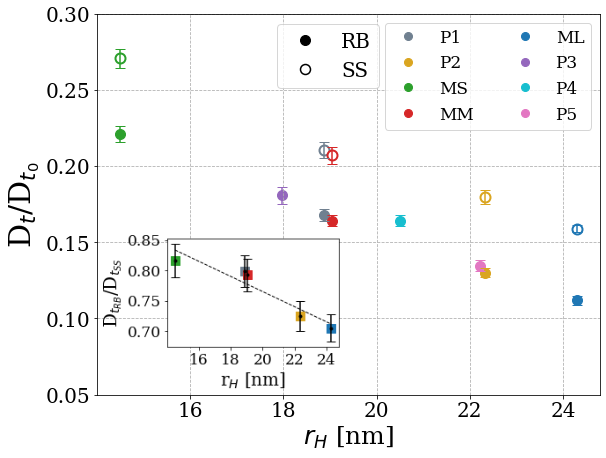}
	\caption{Main: normalized, translational diffusivities for the NPs as function of their hydrodynamic radius ($r_\mathrm{H}$). Filled and empty circles indicate RB and equivalent h-SS representations for systems with same colors, respectively. Inset: diffusivities for the RB models normalized over the equivalent h-SS ones, as function of their hydrodynamic radius $r_\mathrm{H}$. }
	\label{fig:results1}
\end{figure}

% The Stokes-Einstein relation states that under identical viscosity and temperature conditions, particles with the same hydrodynamic size have the same diffusion rate in pure solvent. Larger $r_H$ correspond to slower mobility, and \textit{vice versa}. Based on previous studies, we would expect tracers with the same $r_H$ to diffuse similarly even in macromolecularly crowded environments, at least for spherical NPs like the ones investigated in this study \cite{Ando2010, Balbo2013, Grimaldo2019}. 
Figure~\ref{fig:results1} shows the normalized translational long-time diffusion coefficients as a function of $r_\mathrm{H}$ for 8 RB NP-PCs (see Table~\ref{tab:systems} for details) and 5 equivalent h-SS (P1, P2, MS, MM, ML). Despite the range of compositions and polydispersities of the PC simulated, all RB systems follow the same monotonic decrease of $D_\mathrm{t}/D_\mathrm{t_0}$ as a function of $r_\mathrm{H}$. Notably, both monodisperse (MS, MM, ML) and polydisperse coronas (P1-P5) align along the same trend line. A similar behaviour is also observed for the h-SS NPs, but the trend is shifted to higher values of $D_\mathrm{t}/D_\mathrm{t_0}$. This shows that the explicitly modelled RB systems (solid circles) diffuse consistently slower than the equivalent h-SS approximations (open circles). For the RB and the h-SS models with the same $r_\mathrm{H}$, the ratio between their $D_\mathrm{t}$ decreases linearly with the size of the tracer, as shown in the inset graph of Figure~\ref{fig:results1}. %Specifically, we find that the larger the size of the tracer, the greater is the difference in diffusion between the RB and the SS models.
As our results suggest, the hydrodynamic size, or Stokes radius, alone does not accurately reflect the dynamics of the NP-PC systems, as the type of model employed (RB or h-SS) determines the diffusivity of a NP even if they have the same hydrodynamic size. 
%Under identical conditions of macromolecular crowding, volume fraction, hydrodynamic size, and translational friction coefficient, the dynamic behaviour of the two representations consistently differs, resulting in two parallel curves instead of the expected single one. 
This discrepancy complicates the effective application of existing theoretical \cite{Tokuyama1995, Phillies1987, Langevin1978, Cai2011} and computational \cite{Ando2010, Blanco2017, Grimaldo2019} models, as well as the interpretation of experimental results that derive the hydrodynamic size from the diffusion behaviour \cite{Skora2020, Kwon2020, Kuhn2011, Babayekhorasani2016}. This issue is particularly relevant for NPs with a PC, as the PC thickness is often derived from the diffusion coefficient \cite{Cui2014} . 
%Our data suggest that the two representations would lead to different hydrodynamic radii, even though in pure solvent, the RB and SS models exhibit the same hydrodynamic size. 
Although the h-SS and RB models have the same hydrodynamic size at infinite dilution, they yield different diffusion coefficients under crowded conditions, highlighting that using the computed $D_t$ to infer size can lead to inconsistent results depending on the tracer's morphology (RB or h-SS). It is clear that in order to accurately represent a NP-PC complex by a SS model, the size of the simplified model cannot be based purely on the $r_\mathrm{H}$ of the NP-PC system, so an alternative descriptor must be derived.

\begin{figure}[h]
\centering
  \includegraphics[width=8.1cm]{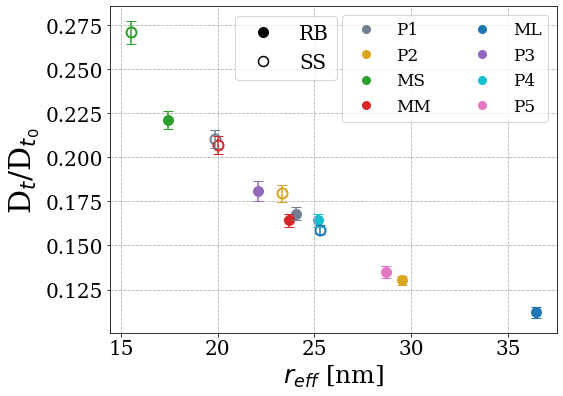}
		\caption{Normalised, translational diffusivities for the NPs as function of their effective radius ($r_\mathrm{eff}$). See text for derivation.}
		\label{fig:results2}
	\end{figure}

Here, we propose the use of the accessible surface area as a geometric descriptor to model the NP-PC SS size in crowding conditions. Our approach is based on the well known Solvent Surface Accessible Area (SASA) \cite{RICHMOND198463}, that corresponds to the area accessible to a solvent molecule, usually water. It is normally calculated by implementing the rolling ball algorithm, which consists in simulating a probe rolling over the surface of interest \cite{Shrake1973} to quantify what surface area is accessible to the probe. The size of the probe can be adjusted depending on the level of detail one wants to include in the representation of the surface. In our case, the crowders will ``see'' the tracer, and depending on their sizes relative to the NP-PC complex, they will perceive more or less details of the surface morphology, affecting the diffusivity of the NP. As in this study the tracer interacts with nanometre-sized crowders, the interest is in deriving a surface accessible to them rather than water molecules as it is commonly done. For the specific medium composition (polydisperse plasma) and volume fraction ($\phi=0.3$) used in our simulations, we find that the optimal size of the probe is $r_\mathrm{probe} \approx 1$ nm. For a more accurate definition and to avoid confusions with the SASA method, we refer to our modified SASA as \textit{Crowders} Accessible Surface Area (CASA), which has been calculated employing the VMD software \cite{Humphrey1996}. Using the calculated CASA, we derive an effective size, $r_\mathrm{eff} = \sqrt{\mathrm{CASA}/4\pi}$ for the RB systems, which will then be used as an alternative to the $r_\mathrm{H}$. Note that for h-SS models $r_\mathrm{eff} = r_\mathrm{H} + r_\mathrm{probe}$, corresponding to the surface accessible to the probe, as setting $r_\mathrm{eff} = r_\mathrm{H}$ would describe the \textit{excluded} surface area instead of the accessible one. 
Figure~\ref{fig:results2} shows the normalized diffusivities plotted against the defined effective radius, $r_\mathrm{eff}$. With this representation, both the detailed RB models and the simplified h-SS models collapse onto a single trend curve. This indicates that $r_\mathrm{eff}$ serves as a meaningful descriptor capable of unifying the diffusional behaviour of different tracer representations.  The P1 RB tracer, for instance, aligns more closely with an h-SS model that has the hydrodynamic radius of the P2 system, rather than that of P1. 

Using a SS approximation based on hydrodynamic size seemed like an obvious choice for modelling diffusion. However, this approach seems to break down under certain conditions. Previous research for proteins suspensions by Balbo \textit{et al}. \cite{Balbo2013} on self-crowded solutions of Bovine Serum Albumin (BSA) and $\gamma$-Globulin (IgG) highlighted the significant role of macromolecular shape in influencing translational diffusion. In their study, treating IgG as a h-SS failed to accurately replicate experimental data, while the spherical assumption worked well for the more globular BSA. Our findings suggest a new key mechanism, particularly relevant for solutions where the tracer (NPs with PC in our case) and the crowders are of different nature. Both the RB and h-SS representations in our study fall under the broad definition of ``globular'' shape, but despite this, their diffusion behaviour differs. We observe that better agreement is achieved when the area of the RB accessible to crowders is taken into account, as systems with similar accessible surface areas (no matter if SS or RB) tend to exhibit comparable diffusion patterns. We now compare our results with those reported by Ando and Skolnick (AS) \cite{Ando2010} on \textit{Escherichia coli} cytoplasm. While a direct comparison of data points is not feasible due to different composition of the protein medium and different volume fractions, a similar distinction between the RB and h-SS representations is evident in AS's study. In their work, the long-time diffusion constants of spherical systems are clearly lower than those of molecular-shaped systems, particularly at higher concentrations. However, AS proposed that macromolecular shape has a minimal impact on diffusion in crowded environments, suggesting that an h-SS is a reasonable approximation for \textit{in vivo} protein diffusion. Specifically, they observed that the difference between molecular-shaped proteins and their spherical approximations was negligible, at least for the experimental diffusion of Green Fluorescent Protein (GFP). Notably, GFP has a $r_\mathrm{H}$ of approximately 2.4 nm, and among the most abundant macromolecules that compose \textit{E. coli} is one of the smallest. We can assume that most crowders in the cytoplasm perceive GFP as a small sphere. To motivate this assumption, we must recall the concept of accessible surface area and how it is computed. In the rolling-ball algorithm, the choice of the probe radius influences the measured surface area, as smaller probes are able to capture finer details, resulting in a larger surface area. Conversely, using a very large probe, maybe larger than the size of the main structure, would ``smooth out'' surface features and approximate the shape as a more uniform, rounded form. The extent to which the smoothed-out object approaches a perfect sphere also depends on its original geometry and overall anisotropy.  However, we can infer that for a small, globular protein such as GFP, the morphology has little to no impact on its motion in a solution made by much larger crowders. This observation implies that the influence of morphology on diffusion, even for globular objects, becomes significant only when the tracer is larger than the crowders. As the size of the tracer increases relative to the crowders, morphological effects become progressively more important in determining diffusion behaviour. Based on our findings and those of AS, the equivalent hydrodynamic sphere model remains effective for small proteins in solution. This is because when the size of the crowders is equal to or larger than the tracer, the surface area accessible for collisions is reduced compared to smaller crowders. For large macromolecular assemblies like NP-PC, the difference between the h-SS approximation and more detailed representations becomes too significant to ignore, and this discrepancy becomes increasingly relevant as the size of the tracer increases with respect of the size of the crowders.
Understanding the influence of morphology on tracer's diffusion is essential for a correct interpretation of \textit{in vivo} conditions, where crowding and heterogeneity are significant \cite{Balbo2013, Skora2020, Grimaldo2019}. From our data, diffusion appears to be significantly influenced by more sophisticated morphological features such as roughness, not just overall shape and hydrodynamic size, with their impact seemingly dependent on the specific composition of the medium, beyond the simple volume fraction of crowders. This makes its interpretation non-straightforward and challenges the assumptions of the single-sphere model for large bodies in polydisperse solutions. Our analysis highlights the contribute of morphology and specifically of roughness as important factor that influences NP-PC mobility under conditions of macromolecular crowding, and we propose the concept of CASA as relevant geometrical descriptor for a correct interpretation of diffusion. In this analysis, the medium composition and volume fraction are kept constant across all simulated tracers, which all exhibit a raspberry-like morphology. Consequently, the size of the probe used in the calculation of $r_\mathrm{eff}$ is also held constant. From our analysis and discussion, the appropriate probe size used to compute the effective surface area is expected to vary with medium composition, concentration, and tracer shape. The optimal value of $r_\mathrm{eff}$ therefore should emerge from the interplay of all these factors. Their combined effects on bridging RB and h-SS diffusion behaviours will be explored in more detail in the following section.

\subsection{Effects of the medium}

\begin{figure}[h]
\centering
  \includegraphics[width=6cm]{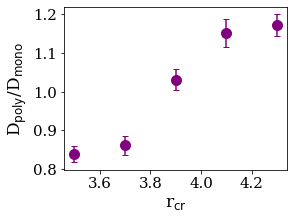}
    \caption{Diffusivity of P1 RB in polydisperse plasma medium normalised over the diffusivities of the same tracer in different mono-crowded media as function of the crowders' size $r_\mathrm{cr}$.}
    \label{fig:poly-mono_media}
\end{figure}

In this section, we focus on the role of the medium on NP diffusion. To do that, we simulated only one NP-PC type (P1) in different media at different concentrations. First, it is important to determine whether the polydisperse plasma medium (with a polydispersity index of $\alpha = 0.26$) can be approximated as a monodisperse suspension ($\alpha = 1$) without affecting the diffusion of the tracer in solution. We derived two effective crowder sizes and we assessed their ability to replicate the behaviour of the polydisperse plasma. To do that, the $D_\mathrm{t}$ of P1 NP in five mono-crowded suspensions was measured, with crowder sizes ($r_\mathrm{cr}$) of $3.5, 3.7, 3.9, 4.1$, and $4.3$ nm, all at a fixed total volume fraction of $\phi = 0.3$. Among these, two sizes were derived as effective representations based on specific criteria, while the remaining three were selected to ensure a representative distribution of data points. Specifically, in the medium with $r_\mathrm{cr} = 3.7$ nm, the size of the crowders is derived as the mean hydrodynamic radius of proteins in the polydisperse plasma, whereas in the medium with $r_\mathrm{cr} = 3.9$ nm, the size of the crowders is derived from colloidal diffusion theory \cite{Grimaldo2019} (see Section 2.2 for details). Figure~\ref{fig:poly-mono_media} shows the $D_\mathrm{t}$ of the P1 RB tracer in polydisperse plasma normalised over the $D_\mathrm{t}$ of the same tracer in the five mono-crowded media ($r_\mathrm{cr}$ of $3.5, 3.7, 3.9, 4.1$, and $4.3$ nm, as detailed above).

Our results show that the diffusivity in $r_\mathrm{cr}=3.9$ nm is the closest match to that of the polydisperse plasma. This suggests that the effective size of crowders, as derived from Reference\cite{Grimaldo2019}, provides a more accurate representation of the crowding effects seen in polydisperse suspensions. However, this agreement could be coincidental, and other scaling approaches might also yield good agreement under the same conditions investigated here. In contrast, the crowder size $r_\mathrm{cr}=3.7$ nm, which was based on the mean protein size in plasma, resulted in a lower diffusivity. Notably, when the effective size is properly chosen, the sensitivity to polydispersity becomes quite low. However, within the narrow range of approximately 3.7–4.1 nm for the effective crowder radius, the $D_\mathrm{poly}/D_\mathrm{mono}$ changes markedly, indicating high sensitivity to $r_\mathrm{cr}$ in this interval. At the extremes of this range, we observe a plateau-like behaviour, suggesting reduced sensitivity to variations in crowder size. Although this simplification may depend on the specific NP-PC simulated and may not be generalizable without incorporating hydrodynamic interactions, it constitutes a step toward establishing a scaling theory for polydisperse solutions, which is currently unavailable. %The present model, which includes volume exclusion interactions, captures key crowding effects in a simplified but physically meaningful way. As also observed in Reference~\cite{Ando2010}, the inclusion of additional interactions such as protein–protein interactions may alter the quantitative results; however, their study shows that the overall diffusive behaviour is mainly described by volume exclusion and hydrodynamic interactions. Furthermore, in the context of simulations, such an approximate medium can aid in simplifying the analysis of diffusion in complex, polydisperse environments.

\begin{figure}[h]
\centering
  \includegraphics[width=8.1cm]{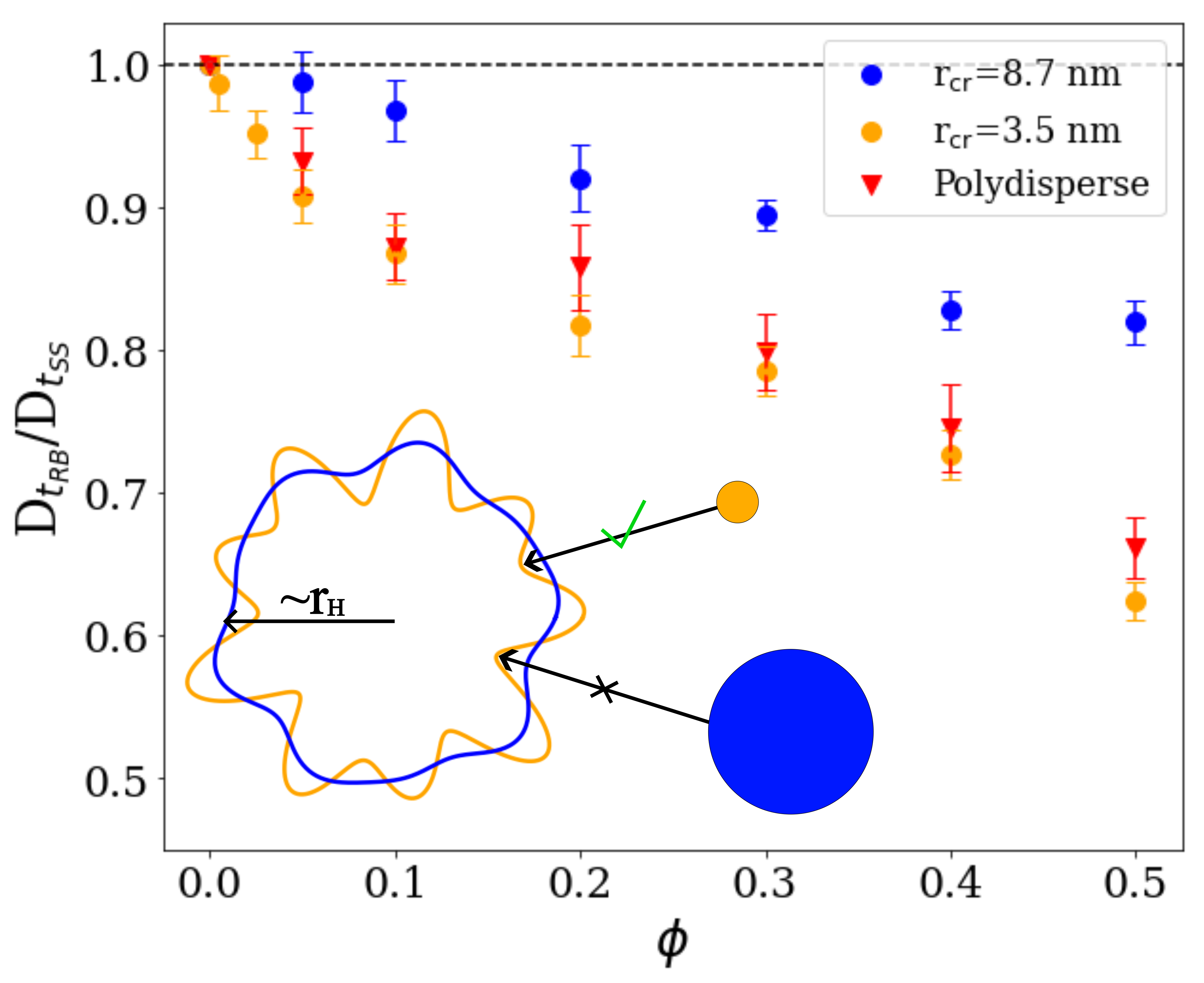}
    \caption{RB diffusivities normalised over the equivalent h-SS ones as function of the volume fraction $\phi$ in polydisperse plasma medium (red triangles), and two mono-crowded media ($r_\mathrm{crowder} = 3.5$ nm in orange circles and $r_\mathrm{crowder} = 8.7$ nm in blue circles). The black dashed line represents the case limit for which $D_{t_\mathrm{RB}}/D_{t_\mathrm{SS}} = 1$. On the bottom left, a schematic representation of the area of the same body accessible to small (orange) and large (blue) crowders, in 1D on the left and 2D on the right. The black, dashed line represents $r_H$, the hydrodynamic size of the body at infinite dilution. The larger is the size of the crowders and the lower is the volume fraction of the solution, the more the accessible area tends towards $r_H$, \textit{i.e.} the crowders in solution ``see'' the tracer as a sphere of equivalent hydrodynamic size.}
    \label{fig:results3}
\end{figure}

Figure~\ref{fig:results3} shows the translational diffusivities of P1 RB normalised over the equivalent h-SS ones ($D_{t_\mathrm{RB}}/D_{t_\mathrm{SS}}$) in three media, two mono-crowded and the polydisperse plasma, at different volume fractions. As the volume fraction of the solution increases, the deviation between the RB representation and the h-SS model becomes more pronounced. We observe a similar trend across all three media. However, the volume fraction where the diffusivities of the two representations become equal seems to depend on the crowders' size. For $r_\mathrm{cr} = 8.7\,\mathrm{nm}$ (blue circles in the figure), they match at $\phi = 0.05$. For $r_\mathrm{cr} = 3.5\,\mathrm{nm}$, the match happens at even lower packing fractions ($0.025$ and $0.005$). These findings are significant for two key reasons: first, they underscore that the influence of tracer's morphology on diffusion increases with the packing fraction of the solution; second, they emphasize that the h-SS approximation is only accurate at very low concentrations, with the specific threshold depending on the size of the crowders. This last statement is of particular interest, as the hydrodynamic radius is widely adopted in colloid-theory based models for the diffusion of macromolecules even at high packing fractions \cite{Ando2010, Balbo2013, Grimaldo2019}. These results, together with the ones shown in Figure~\ref{fig:results2}, would suggest that the size ratio between the tracer and the crowder, together with the tracer's morphology and medium volume fraction, are key factors in the correct interpretation of macromolecular diffusion. When a tracer interacts with a population of much smaller crowders, the surface area available for collisions becomes bigger compared to interactions with larger crowders, as can be depicted from the 2D representation in Figure~\ref{fig:results3}. This is particularly significant in the case of patchy-like bodies, which may feature concave regions that can only be accessed by smaller crowders in the solution. 
%Moreover, considering that a single larger crowder occupies the same volume as multiple smaller ones, the frequency of collisions experienced by the tracer from the larger crowder will be lower than the cumulative effect of the numerous smaller crowders. The increased number of volume-excluded interactions with the immediate neighbours leads to a reduction in overall mobility. In crowded biological environments, where there is a wide distribution of molecular sizes, the prevalence of interactions with smaller crowders, driven by their higher collision rates due to the access to larger surface areas and higher mobility, enhances the random fluctuations in the tracer's position, resulting in a reduced diffusion. This underscores the influence of morphology in the interactions between a tracer and surrounding crowders, and highlights the limitations of a straightforward application of colloid theory to large tracers in polydisperse mixtures. 
In this discussion, we assumed that the morphological properties of the analysed systems remain unchanged across different concentrations. It is known that macromolecular crowding induces changes in the folding and compactness of individual proteins \cite{Miller2016, Nguemaha2018, Haas-Neill2024, Tokuriki2004}, but its impact on more complex structures, such as the PC, remains underexplored. It is unclear whether macromolecular crowding influences morphology, surface roughness, and overall structural integrity of the PC in a way similar to individual proteins. However, a recent study on the biomolecular corona surrounding vesicles hypothesizes that discrete regions of protruding aggregates likely form the corona’s architecture, even at high concentrations \cite{Bergese2025}. This finding supports the notion of an anisotropic, patchy layer of biomolecules around the central particle, even under macromolecular crowding conditions, aligning with the model proposed here. 

\begin{figure}[h]
\centering
  \includegraphics[width=6cm]{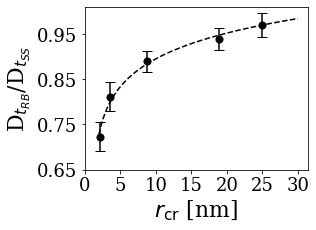}
	\caption{RB diffusivities normalised over the equivalent h-SS ones as function of $r_\mathrm{cr}/r_\mathrm{tr}$. Line is only to guide the eye.}
	\label{fig:results4}
\end{figure}

From the inset in Figure~\ref{fig:results1}, we already observed that $D_\mathrm{t_{RB}}/D_\mathrm{t_{SS}}$ decreases linearly as the size of the tracer $r_\mathrm{tr}$ increases. Figure~\ref{fig:results3} further proves that at same volume fraction points, larger crowders in the medium result in higher values of $D_\mathrm{t_{RB}}/D_\mathrm{t_{SS}}$. These observations would suggest that the choice of the model (RB or h-SS) depends on the size ratio between tracer and crowders, and that the h-SS approximation may be valid within certain limits.
To explore this hypothesis, we analysed translational diffusivities of the P1 RB and h-SS models (with constant tracer radius $r_\mathrm{tr}$ = 18.9 nm) in five mono-crowded environments, each with different crowder sizes ($r_\mathrm{cr}=2.1,3.5,8.7,18.9$ and 25 nm), while keeping the total volume fraction constant at $\phi = 0.3$. As shown in Figure~\ref{fig:results4}, normalised diffusivity strongly depends on the crowder-to-tracer size ratio. Specifically, the diffusion coefficients of RB and h-SS tracers tend to converge as the ratio $r_\mathrm{cr}/r_\mathrm{tr}$ increases. This indicates that, for the specific raspberry shape analysed in this study, when crowders are larger than the tracer, the tracer morphology has a negligible impact on diffusion. We compare now these findings with the results of Ando and Skolnick~\cite{Ando2010} and Balbo \textit{et al.}~\cite{Balbo2013}. In the former study, a h-SS representation was sufficient to capture the diffusion of GFP in a complex, polydisperse cytoplasmic environment, emphasizing the dominant role of hydrodynamic interactions, which were treated with a detailed model that included far-field, many-body and near-field hydrodynamic interactions. In contrast, Balbo \textit{et al.} found that the same h-SS approximation failed to accurately describe the diffusion of IgG under self-crowding conditions, while it worked well for BSA under identical conditions. They concluded that molecular shape could have a stronger impact on diffusion than other factors. Since Balbo and Ando treat HIs using different approximations, their results are not directly comparable to each other, nor to ours, as we do not include HIs at all in this model. Our focus was on morphological effects, and excluding other interactions from the model allowed us to isolate their influence; the present results offer a possible interpretation along these lines. Our results suggest that the influence of tracer morphology on diffusion becomes negligible when the tracer is significantly smaller than the surrounding crowders, as in the case of GFP in the cytoplasmic solution. However, when the tracer and crowders are of comparable size, as in the self-crowding conditions for BSA and IgG, shape effects become more pronounced due to volume exclusion interactions. The fact that the h-SS model succeeded for BSA but failed for IgG under the same size ratio conditions (i.e., $r_\mathrm{tr}/r_\mathrm{cr} = 1$) highlights the role of the tracer intrinsic shape. BSA, being globular, is well approximated by a sphere, while IgG, with its extended Y-shaped structure, deviates substantially from spherical symmetry and shows model-dependent diffusion behaviour at the $r_\mathrm{tr}/r_\mathrm{cr} = 1$. This highlights that the validity of the h-SS model is not governed solely by the overall shape (i.e: globular, Y-shaped, rod-like), but also by the tracer-to-crowder size ratio. Therefore, the interplay between these two factors must be jointly considered when choosing an appropriate model to describe diffusion in crowded environments.

	\begin{figure}[h]
    \centering
	\includegraphics[width=8.1cm]{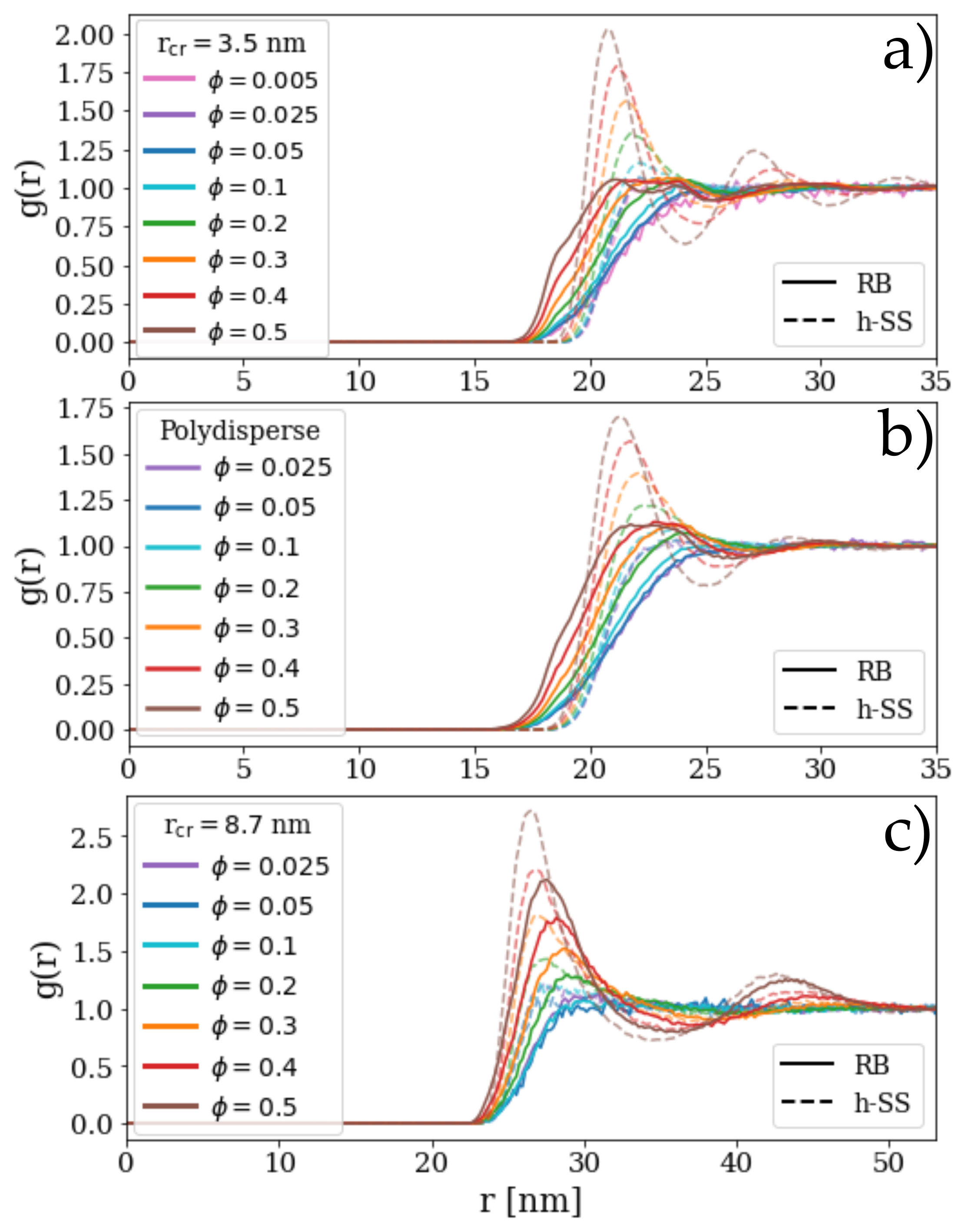}
		\caption{Radial distribution function of P1 NP in a) monodisperse medium  with $r_\mathrm{cr}=3.5$ nm at 8 different volume fraction b) polydisperse plasma medium at 7 different volume fraction, c) monodisperse medium  with $r_\mathrm{cr}=8.7$ nm at 7 different volume fraction. In all plots, solid line indicates the RB model, dashed line indicates h-SS model.}
		\label{fig:rdf media}
	\end{figure}

Figure~\ref{fig:rdf media} shows the Radial Distribution Function (RDF) computed for the same systems as in Figure~\ref{fig:results3}, using the center-of-mass distance between the NP-PC complex and crowders. The RDFs, resolved at 0.02 with binning adjusted for system size, highlight how crowder size, polydispersity, and tracer geometry shape local structure. All systems exhibit increased structuring with crowding, with the h-SS model showing sharper peaks due to its smooth, spherical geometry, indicating stronger local ordering. In contrast, the RB model, comprising multiple beads, presents broader, less defined peaks, reflecting a more irregular surface and reduced local ordering.

Crowder size further modulates structuring: smaller crowders (Figure~\ref{fig:rdf media}-a) produce broader RDFs, particularly for the RB model, while larger crowders (Figure~\ref{fig:rdf media}-c) yield sharper peaks and reduced sensitivity to tracer geometry. Polydispersity (Figure~\ref{fig:rdf media}-b) smooths RDF profiles, reflecting heterogeneous environments. As crowder size increases relative to the tracer, the RDFs and diffusion coefficients of the two models converge. However, at higher volume fractions, differences between h-SS and RB models grow more pronounced, emphasizing the role of tracer morphology under crowded conditions.

Although translational diffusion analysis showed that the RB model can be approximated by an isotropic representation, provided it accounts for both hydrodynamic size and morphological contributions (dependent on tracer/crowder size ratio and volume fraction), this simplification sacrifices detail on local structuring. Representing a NP-PC as an equivalent sphere may suffice for translational diffusion analysis, but it limits insight into microscale interactions, a critical factor for future studies aiming to capture both translational dynamics and spatial organization.

\section*{Conclusions and outlook}
In this study, we presented a mesoscale model for NPs with their characteristic PCs in crowded media. We investigated the diffusion of tracers significantly larger than the crowding agents in solution. This focus on large tracer diffusion in crowded, polydisperse environments appears under-explored in the current literature, and the findings could provide valuable insights, particularly regarding NP diffusion. Here the PC is treated as rigid-body, under the assumption that, once formed, it is steadily adsorbed on the NP within the time scales relevant for the diffusion. Proteins, both in solution and in the PC, have been modelled as soft repulsive spheres of equivalent hydrodynamic radii and interacting via volume exclusion interactions. This approximation allowed us to examine how specific properties of the tracer and of the medium influenced translational diffusion. Different compositions of the PC and different sized crowding agents have been tested, at several volume fractions. 

In the analysis of the diffusivity in the long time, the combined effects of macromolecular crowding and morphology emerge as critical factors for the derivation of the tracer's effective size and, consequently, of the PC's thickness. We find that as the volume fraction of the crowding agents increases, the diffusion of the molecularly-shaped representations diverges significantly from the equivalent single-sphere ones. This suggests that deriving the overall size/PC thickness from the diffusion coefficient using the Stokes-Einstein relationship at volume fractions $> 0.05$ can lead to incorrect estimations. The fact that globular bodies with the same hydrodynamic radius at infinite dilution exhibit significantly different diffusion behaviours at higher packing fractions challenges the accuracy of size estimations under crowded conditions, calling into question the validity of current approaches for determining true sizes in such environments. 

Our results show that the deviation from the hydrodynamic single-sphere prediction scales linearly with both the tracer-to-crowder size ratio and the volume fraction. This scaling reveals that, in polydisperse systems containing macromolecules of widely differing sizes, the hydrodynamic single-sphere approximation fails to reliably describe the diffusion of larger macromolecules, even though it remains adequate for smaller ones. This finding provides a framework for reconciling previous discrepancies between experimental or simulation results and the hydrodynamic single-sphere model, especially for more complex objects in crowded environments. It also underscores that tracer morphology and the relative size ratio between tracer and crowders are key determinants of diffusion.

We therefore propose the accessible surface area as a central parameter in diffusion analysis and recommend its determination for both experimental and computational studies. The probe size required for accurately calculating this accessible surface area (and for meaningfully applying the single-sphere approximation) appears to be specific to the system under investigation. That different structural models of the same tracer, under identical crowding conditions, yielded significantly different diffusion coefficients directly demonstrates the decisive role of morphology. Our results suggest that this morphological effect arises from two main factors: 1) smaller crowders have access to more surface area of the NP-PC complex; 2) greater volume fractions increase the sensitivity of diffusion to the specific morphology of the tracer. Consequently, the diffusivity of the NP–PC complex is jointly determined by the overall volume occupation of the crowders, their size distribution relative to the tracer, and the tracer’s morphology, factors that must be considered when extending single-sphere models to more complex crowding scenarios.

It is important to note that this study does not account for hydrodynamic interactions or the shape of the crowders. Both factors are crucial for fully formulating and deriving a scaling law, as well as for enabling a direct comparison between simulations and experimental data \cite{Ando2010, Skora2020}. Future work should prioritize incorporating these aspects into the proposed model. Furthermore, a detailed assessment of the effects of macromolecular crowding on the features of the PC is needed. It remains unclear whether the raspberry-like rigid model is still valid at high packing fractions and how high concentrations of crowders might affect PC's organization, thickness, softness and morphology. Despite this, our findings provide valuable insights into NP diffusion in polydisperse, concentrated media, improving our understanding of NP mobility in biological environments, that is crucial for predicting the toxicological and pharmacokinetic behaviour of these nanomaterials in therapeutic and nanomedical applications. Furthermore, these results could be extended to other systems, such as large proteins or molecular assemblies, to advance the study of macromolecular diffusion of large, anisotropic bodies in polydisperse solutions.

%\section*{Author contributions}
%We strongly encourage authors to include author contributions and recommend using \href{https://casrai.org/credit/}{CRediT} for standardised contribution descriptions. Please refer to our general \href{https://www.rsc.org/journals-books-databases/journal-authors-reviewers/author-responsibilities/}{author guidelines} for more information about authorship.

\section*{Conflicts of interest}
There are no conflicts to declare.

\section*{Data availability}

The data supporting this article have been included as part of the Appendix.

\section*{Acknowledgements}
The research conducted in this publication was funded by the Irish Research Council under grant number GOIPG/2023/5034 and by the TU Dublin First Time Supervisor Award.
\\
The authors wish to acknowledge the Irish Centre for High-End Computing (ICHEC) for the provision of computational facilities and support.

%%% Uncomment this line and comment out the ``thebibliography'' section below to use the external .bib file (using bibtex) .
\bibliographystyle{plain}
\bibliography{references} 





% \end{thebibliography}

% -----------------------------------------------
% Supporting Information Section (one-column)
% -----------------------------------------------
\clearpage           % ensure it starts on a new page
%\onecolumn           % switch to one-column layout

%\setcounter{maintext}{0}
\setcounter{figure}{0}
\setcounter{table}{0}
\setcounter{equation}{0}
\renewcommand{\thefigure}{S\arabic{figure}}
\renewcommand{\thetable}{S\arabic{table}}
\renewcommand{\theequation}{S\arabic{equation}}

\section*{Appendix}

Additional computational details, figures, and tables are provided below.

\subsection*{Derivation of Moment of Inertia for rigid bodies}
In simulations of diffusion involving rigid bodies, the moment of inertia (MOI) must be derived for accurately modelling rotational dynamics. The MOI quantifies an object's resistance to rotational motion, influenced by its mass distribution. For simple shapes like a pendulum, the moment of inertia is straightforwardly calculated as $I = mr^2$, where $m$ is mass and $r$ is distance from the rotation axis. For complex objects, it's computed as $\sum_{i} m_ir_i^2$, accounting for individual particles. The moment of inertia is axis-dependent, described comprehensively by the inertia tensor $\mathbf{I}$, with diagonal elements representing principal moments of inertia. These can be obtained by diagonalizing $\mathbf{I}$, finding eigenvalues and eigenvectors $\lambda_i$ and $\mathbf{k}_i$. The rotation matrix $\mathbf{P}$ is constructed from eigenvectors. For any rigid body and any point in it there are three orthogonal principal axes for which the matrix representing the inertia tensor is diagonal. By diagonalizing the inertia tensor, the coordinate system is aligned with these principal axes, highlighting the $principal$ $moments$, that correspond to the elements of the diagonalized matrix. This can be obtained by applying the rotation to the inertia tensor itself, as
	\begin{equation}
		\begin{aligned}
			\mathbf{I}' &= \mathbf{PIP}^T = \begin{pmatrix}
				\mathbf{k}_1^T \\
				\mathbf{k}_2^T \\
				\mathbf{k}_3^T
			\end{pmatrix}\mathbf{I} \begin{pmatrix}
				\mathbf{k}_1 & \mathbf{k}_2 & \mathbf{k}_3
			\end{pmatrix} = \\ &= \begin{pmatrix}
				\mathbf{k}_1^T \\
				\mathbf{k}_2^T \\
				\mathbf{k}_3^T 
			\end{pmatrix} \begin{pmatrix}
				\lambda_1\mathbf{k}_1 & \lambda_2\mathbf{k}_2 & \lambda_3\mathbf{k}_3
			\end{pmatrix} = \begin{pmatrix}
				\lambda_1 & 0 & 0 \\
				0 & \lambda_2 & 0 \\
				0 & 0 & \lambda_3
			\end{pmatrix}.
		\end{aligned}
	\end{equation}
	
In this way, the inertia tensor is transformed from the original coordinate system to a new one where the inertia tensor is aligned with the principal axes.

Using the aligned structures, we obtained the hydrodynamic properties at infinite dilution, such as the $D_{t_0}$ and the $3\times3$ rotational diffusion matrix, with the HYDRO++ software. These data were then used to define the drag coefficient and rotational drag coefficient tensor in the Brownian Dynamics integrator.

The total mass and moment of inertia of the rigid body are determined by the mass and moment of inertia of the central particle, while the masses of the constituent particles are not taken into account. The central particle is positioned at the centre of mass of the rigid body, and its orientation quaternion specifies the rotation from a reference frame in simulation box. After defining a rigid body, the positions and orientations of the constituent particles are set relative to the position and orientation of the central particles. Forces, energies and torques are transferred from the constituent particles to the central one, and they are added all together and used to integrate the equations of motion of the central particle, that should represent the entire rigid body.

\subsection*{Data analysis}
1. \textbf{Average Diffusion Coefficient} \\
\[
\bar{D} = \frac{1}{N} \sum_{i=1}^{N} D_i
\]
$N$ = number of replica for each simulated system.

2. \textbf{Standard Error of the Diffusion Coefficient} \\
\[
\sigma_{\bar{D}} = \frac{\sigma_{D}}{\sqrt{N}}
\]
where
\[
\sigma_{D} = \sqrt{\frac{1}{N-1} \sum_{i=1}^{N} \left( D_i - \bar{D} \right)^2}
\]

3. \textbf{Propagation of Error \cite{libretexts_propagation_of_error} for the RB/SS ratio} \\
\[
R = \frac{\bar{D}_{RB}}{\bar{D}_{SS}}
\]
\[
\sigma_R = R \cdot \sqrt{\left( \frac{\sigma_{\bar{D}_{RB}}}{\bar{D}_{RB}} \right)^2 + \left( \frac{\sigma_{\bar{D}_{SS}}}{\bar{D}_{SS}} \right)^2}
\]

\onecolumn

\subsection*{Plasma and corona proteins' details}
\begin{center}
\begin{table}[H]
		\centering
		\rotatebox{90}{%
		\begin{tabular}{ccccccc}
			\toprule
			Name & Structure ID & Molarity [M] & MW [kDa] & r [nm] & D$_{t_0}$ [nm\textsuperscript{2}/$\mu$s] & D$_{r_0}$ [$\mu$s\textsuperscript{-1}] \\
			\midrule
			Albumin & 1AO6 & 650 & 66.6 & 3.5 & 61.3 & 3.6 \\
			Immunoglobulin G & 1HZH & 70 & 151.6 & 5.5 & 39.1 & 0.9 \\
			Apolipoprotein A-I & 1AV1 & 54 & 93.8 & 4.9 & 44.1 & 1.2 \\
			Apolipoprotein A-II & P08519 & 41 & 11.2 & 2.4 & 89.8 & 9.4 \\
			Transferrin & 1D3K & 35 & 36.5 & 2.6 & 81.4 & 8.5 \\
			$\alpha$1-Proteinase inhibitor & 1ATU & 29 & 42.0 & 2.9 & 74.7 & 6.6 \\
			Transthyretin & 4TLT & 23 & 13.7 & 2.4 & 89.2 & 11.2 \\
			$\alpha$1-Acid glycoprotein & 3APU & 21 & 45.8 & 2.3 & 93.1 & 14.0 \\
			Hemopexin & P02790 & 15 & 51.7 & 3.9 & 55.3 & 2.5 \\
			Immunoglobulin A & 1IGA & 14 & 148.6 & 6.2 & 34.9 & 0.8 \\
			Apolipoprotein C-III & 2JQ3 & 13 & 8.8 & 2.3 & 92.4 & 10.7 \\
			$\alpha$2-Macroglobulin & 7VON & 12 & 160.1 & 5.0 & 42.7 & 1.2 \\
			$\alpha$2-HS glycoprotein & P02765 & 12 & 39.3 & 3.9 & 55.5 & 2.3 \\
			Hepatoglobin & P00738 & 12 & 45.2 & 3.7 & 57.5 & 2.7 \\
			Gc globulin & 1J78 & 11 & 51.9 & 3.3 & 65.7 & 4.4 \\
			Apolipoprotein C-I & P02654 & 9 & 9.3 & 2.1 & 102.4 & 15.8 \\
			Fibrinogen & 3GHG-BA1 & 9 & 325.7 & 8.7 & 24.6 & 0.5 \\
			Complement C3 & 2A73 & 8 & 185.7 & 5.2 & 41.0 & 1.1 \\
			$\alpha$1-Antichymotrypsin & 1AS4 & 7 & 42.9 & 2.8 & 75.8 & 7.4 \\
			$\beta$2-Glycoprotein I & 6V06 & 6 & 42.6 & 4.4 & 49.1 & 2.1 \\
			Vitronectin & P04004 & 6 & 54.3 & 4.3 & 50.2 & 1.9 \\
			$\alpha$1-B Glycoprotein & P04217 & 5 & 54.3 & 3.8 & 56.7 & 2.7 \\
			Apolipoprotein A-IV & 3S84 & 5 & 63.3 & 4.0 & 53.1 & 2.9 \\
			Apolipoprotein C-II & P02655 & 5 & 11.3 & 2.3 & 93.4 & 12.23 \\
			Ceruloplasmin & 4ENZ & 3 & 123.4 & 4.0 & 54.3 & 2.6 \\
			Antithrombin III & 2B4X & 3 & 99.3 & 4.0 & 53.1 & 2.4 \\
			Inter $\alpha$-trypsin inhibitor & 6FPZ & 3 & 74.3 & 3.5 & 62.1 & 3.8 \\
			Plasminogen & 4DUU & 3 & 88.5 & 3.8 & 56.8 & 2.9 \\
			Retinol-binding protein & 1RBP & 2 & 21.3 & 2.2 & 95.7 & 14.3 \\
			\bottomrule
		\end{tabular}
		}
		\caption{Plasma Proteins Data. The 4-character alphanumeric identifiers indicate structures obtained from the Protein Data Bank, while the structures obtained from Alphafold are indicated with the letter P followed by five numbers.}
		\label{tab:plasma_proteins}
	\end{table}
\end{center}
\vspace*{\fill}    

\begin{table}[H]
	\centering
	\caption{List of proteins characterizing the corona in the P1 NP, with experimental relative abundance obtained from \cite{Lai2017} and size derived from the full atomistic structures.}
	\label{tab:Ag_corona}
	\begin{tabular}{ccccccc}
		\toprule
		Protein & Relative abundance [$\%$]& Stokes radius [nm] \\
		\midrule
		Kininogen  & 27.6 & 5.5 \\
		Apolipoprotein E & 12.4 & 2.6 \\
		Vitronectin & 12.3 & 4.3 \\
		Plasma serine protease inhibitor & 5.4 & 4.9 \\
		Fibrinogen & 8.4 & 8.7 \\
		Coagulation factor V & 4.2 & 4.8 \\
		Isoform 2 of plasma protease C1 inhibitor & 3.8 & 4.5 \\
		Plasma kallikrein & 3.7 & 3.4 \\
    	Coagulation factor XI & 2.6 & 3.5 \\
		Histidine-rich glycoprotein & 2.4 & 4.7 \\
		Immunoglobulin G & 1.8 & 5.5 \\
		Complement C3 & 1.6 & 5.2 \\
		\bottomrule
	\end{tabular}
\end{table}	

	\begin{table}[H]
	\centering
	\caption{List of proteins characterizing the corona in the P2 NP, with experimental relative abundance obtained from \cite{Lai2017} and size derived from the full atomistic structures.}
	\label{tab:Au_corona}
	\begin{tabular}{ccccccc}
	\toprule
	Protein & Relative Abundance [$\%$] & Stokes radius [nm] \\
	\midrule
		Fibrinogen & 41.1 & 8.7 \\
		ITIH4 protein & 16.1 & 5.4 \\
		Kininogen & 7.6 & 5.5 \\
		Complement C3 & 6.3 & 5.2 \\
		Plasma serine protease inhibitor & 3.7 & 4.9 \\
		Isoform E of proteoglycan 4 & 3.0 & 7.4 \\
		Vitronectin & 3.0 & 4.3 \\
		Coagulation factor V & 1.6 & 4.8 \\
		Apolipoprotein E & 1.6 & 2.6 \\
		Coagulation factor XI & 1.0 & 3.5 \\
		Carboxypeptidase & 1.0 & 2.8 \\
		\bottomrule
	\end{tabular}
    \end{table}

\begin{figure}[h]
    \centering
    % First image
    \begin{minipage}[t]{0.45\textwidth}
        \centering
        \includegraphics[width=\textwidth]{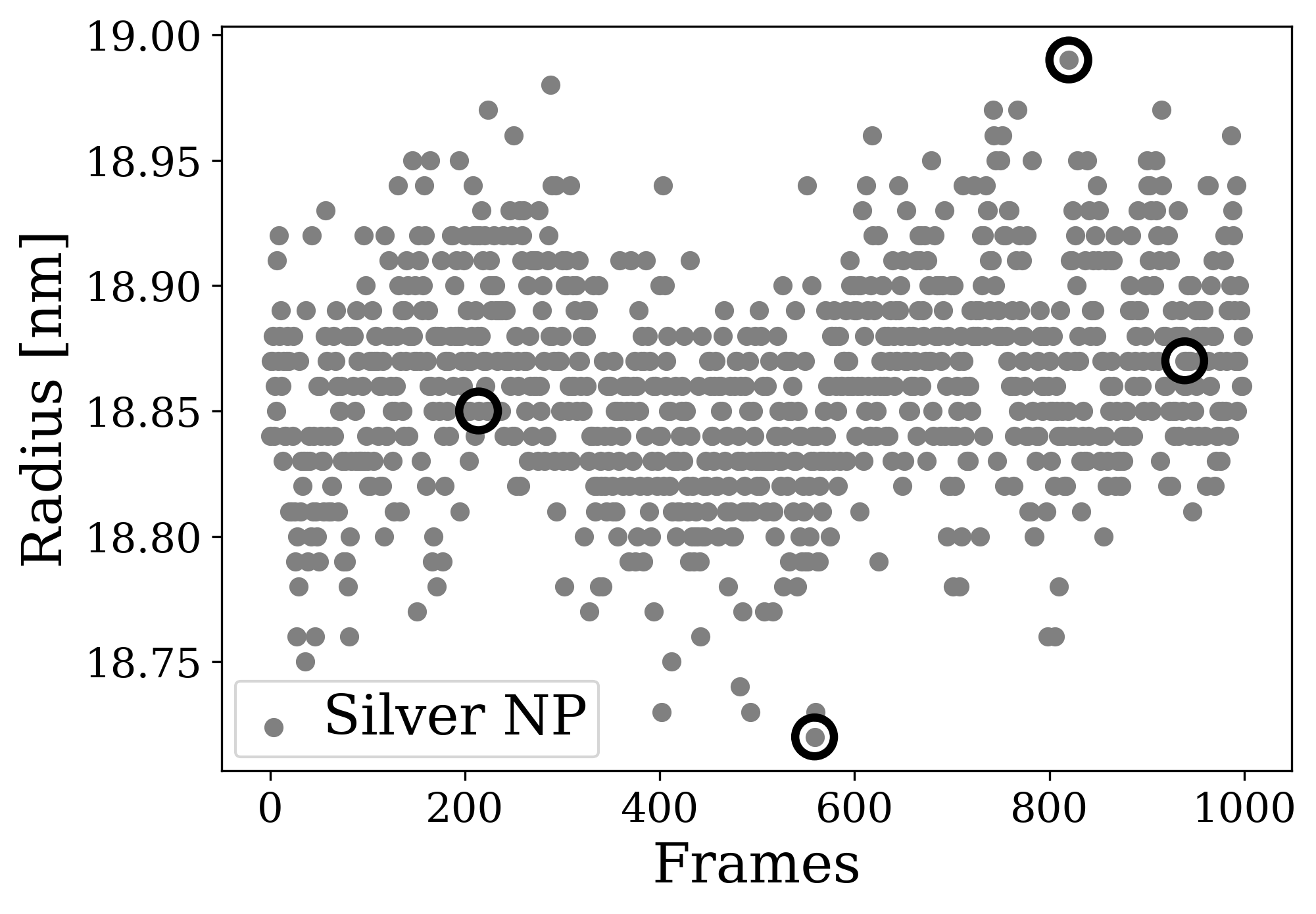}
        \caption{Hydrodynamic sizes of 1000 different morphologies of the PC in the P1 NP. The black circles indicate the morphologies selected in this study. The experimental molarity of each protein is obtained from Ref.\cite{Hortin2006}}
        \label{fig:img1}
    \end{minipage}
    \hfill
    % Second image
    \begin{minipage}[t]{0.45\textwidth}
        \centering
        \includegraphics[width=\textwidth]{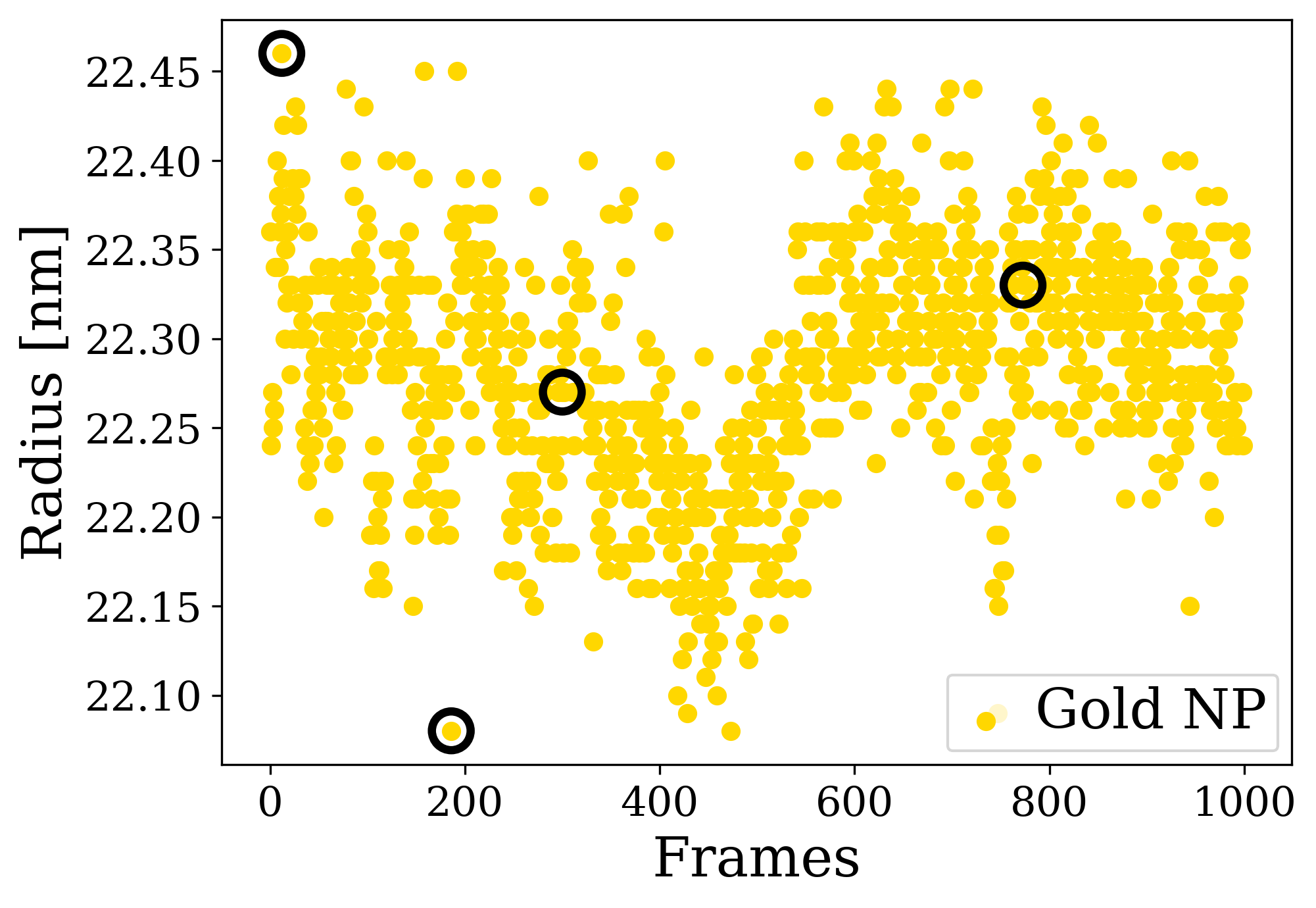}
        \caption{Hydrodynamic sizes of 1000 different morphologies of the PC in the P2 NP. The black circles indicate the morphologies selected in this study.}
        \label{fig:img2}
    \end{minipage}
\end{figure}

\twocolumn
\subsection*{Mean Squared Displacement}

The log-log plots of the MSD vs. time were analysed and the linear fits of the MSD vs. time curves were performed for the long (i.e. ``diffusive'') regimes. The particles are considered in the diffusive regime when the slope from a log-log plot of the MSD is approximately 1 \cite{Maginn2020}. From the linear fit with the MSD in long-time, normal, diffusive regime, the translational diffusion coefficient is obtained as 
\begin{equation}
    D_t = \frac{m}{6}
\end{equation}
with $m$ slope of the linear fit.

\begin{figure}[H]
		\centering
			\centering
			\includegraphics[width=0.4\textwidth]{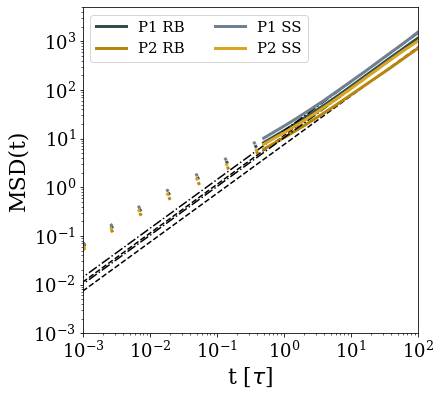}
		\caption{Ensemble-averaged, center-of-mass MSD(t). Black dashdot/dashed lines correspond to fits to the long-time regime}
		\label{fig:fit1}
	\end{figure}

 \begin{figure}[H]
		\centering
			\centering
			\includegraphics[width=0.4\textwidth]{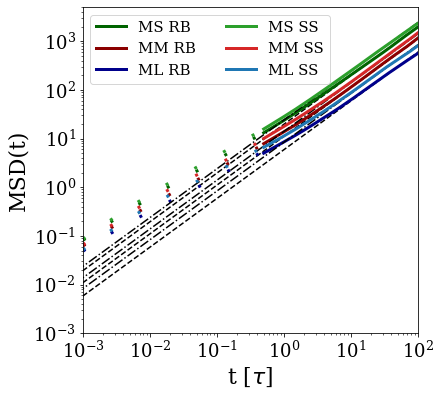}
		\caption{Ensemble-averaged, center-of-mass MSD(t). Black dashdot/dashed lines correspond to fits to the long-time regime}
		\label{fig:fit2}
	\end{figure}

  \begin{figure}[H]
		\centering
			\centering
			\includegraphics[width=0.4\textwidth]{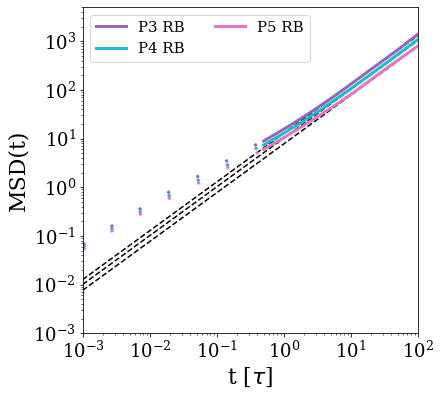}
		\caption{Ensemble-averaged, center-of-mass MSD(t). Black dashdot/dashed lines correspond to fits to the long-time regime}
		\label{fig:fit3}
	\end{figure}

  \begin{figure}[H]
		\centering
			\centering
			\includegraphics[width=0.5\textwidth]{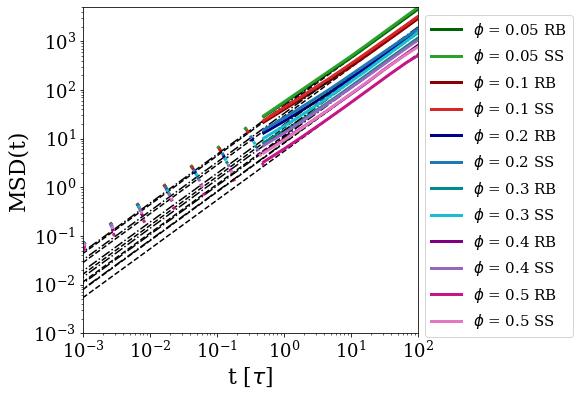}
		\caption{Ensemble-averaged, center-of-mass MSD(t) in Medium $\#$0 (polydisperse plasma). Black dashdot/dashed lines correspond to fits to the long-time regime}
		\label{fig:fit4}
	\end{figure}

  \begin{figure}[H]
		\centering
			\centering
			\includegraphics[width=0.5\textwidth]{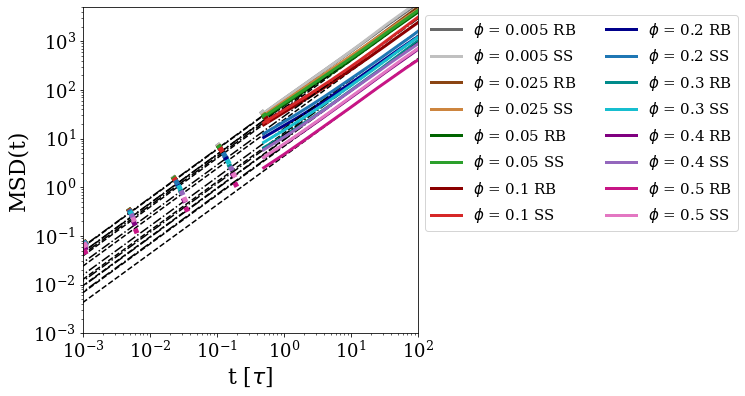}
		\caption{Ensemble-averaged, center-of-mass MSD(t) in Medium $\#$1 (one type of crowder). Black dashdot/dashed lines correspond to fits to the long-time regime}
		\label{fig:fit5}
	\end{figure}

  \begin{figure}[H]
		\centering
			\centering
			\includegraphics[width=0.5\textwidth]{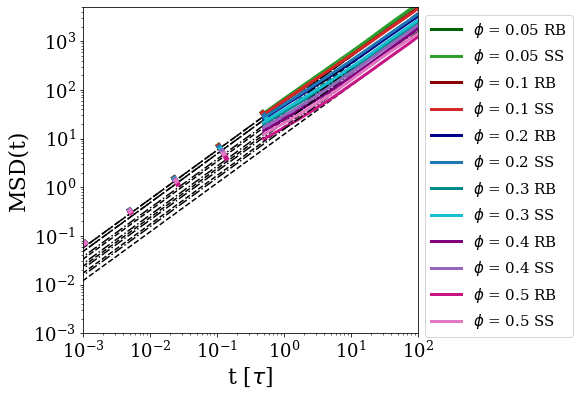}
		\caption{Ensemble-averaged, center-of-mass MSD(t) in Medium $\#$5 (one type of crowder). Black dashdot/dashed lines correspond to fits to the long-time regime}
		\label{fig:fit6}
	\end{figure}

%\bibliographystyle{IEEEtran}
%\bibliography{bibliography}

\end{document}